\documentclass[journal=jacsat,manuscript=communication]{achemso}
\setkeys{acs}{articletitle=true}
\usepackage{amsmath,amssymb,amsthm}
\usepackage{graphicx}
\usepackage{booktabs}
\usepackage{multirow}
\usepackage{array}
\usepackage{mhchem}
\usepackage{algorithm}
\usepackage{algpseudocode}
\usepackage{xcolor}
\usepackage{verbatim}
\usepackage{tikz}
\usepackage[normalem]{ulem}
\usetikzlibrary{positioning,arrows.meta}
\usepackage{titlesec}
\titleformat{\section}{\normalfont\large\bfseries}{\thesection}{1em}{}
\titleformat{\subsection}{\normalfont\normalsize\bfseries}{\thesubsection}{1em}{}
\titleformat{\subsubsection}{\normalfont\normalsize\bfseries}{\thesubsubsection}{1em}{}

\newtheorem{remark}{Remark}
\newtheorem{lemma}{Lemma}

\SectionNumbersOn

\author{Zongwei Huang}
\affiliation[University of Michigan]
  {Department of Chemistry, University of Michigan, Ann Arbor, Michigan 48109, United States}

\author{Bradley S Harris}
\affiliation{Physical Sciences Division, Pacific Northwest National Laboratory, Richland, Washington 99352, United States}

\author{Slater T. Bakenhaster}
\affiliation{Department of Chemistry and Biochemistry,
Kent State University,
Kent, Ohio 44242, United States}

\author{Roshan Khatri}
\affiliation{Department of Chemistry and Biochemistry,
Kent State University,
Kent, Ohio 44242, United States}

\author{Margaret S. Cheung}
\affiliation{Physical Sciences Division, Pacific Northwest National Laboratory, Richland, Washington 99352, United States}

\author{Barry D. Dunietz}
\affiliation{Department of Chemistry and Biochemistry,
Kent State University,
Kent, Ohio 44242, United States}
\email{bdunietz@ksu.edu}

\author{Marcel D. Baer}
\email{marcel.baer@pnnl.gov}
\affiliation{Physical Sciences Division, Pacific Northwest National Laboratory, Richland, Washington 99352, United States}

\author{Eitan Geva}
\email{eitan@umich.edu}
\affiliation[University of Michigan]
  {Department of Chemistry, University of Michigan, Ann Arbor, Michigan 48109, United States}

\title{
Synthetic Stochastic Time Series from Generative Models
}


\begin{document}

\begin{abstract}
Obtaining statistically converged time-series data
represents a
fundamental challenge in computational chemistry, where sampling is
constrained by the prohibitive computational cost of high-fidelity simulations.
The accuracy attainable from a finite simulation budget is governed by the
standard-error scaling $1/\sqrt{n}$, rendering exhaustive sampling challenging for
systems of realistic complexity.
Here we show that generative models trained on high-quality stochastic
trajectories can reproduce key statistical properties of the underlying
process and generate additional synthetic trajectories at substantially
lower computational cost, thereby enhancing the sampling accessible from a
finite set of simulations.
We present a
systematic comparison of four generative architectures across two complementary
data sources: an analytically tractable Ornstein--Uhlenbeck reference and
donor--acceptor energy-gap trajectories from all-atom molecular
dynamics simulations.
Among the four architectures tested, the denoising diffusion probabilistic model reproduces the target statistics most consistently, across both data sources and across the range of correlation times examined, and generates far faster than the sequential Transformer.
Specifically, for the energy-gap trajectories, augmentation from 0.1 ns of training data yields statistical estimates comparable in accuracy to those obtained from 0.25--0.50 ns of direct MD sampling.
We further show that the benefit of synthetic augmentation is bounded by the information contained in the training data, producing a quality floor determined by both the training data and the generative model.
Our findings establish diffusion-based generative models as an effective route to data-driven augmentation of stochastic time series in computational chemistry.
\end{abstract}

\begin{figure}[t]
  \centering
  \includegraphics[width=0.5\textwidth]{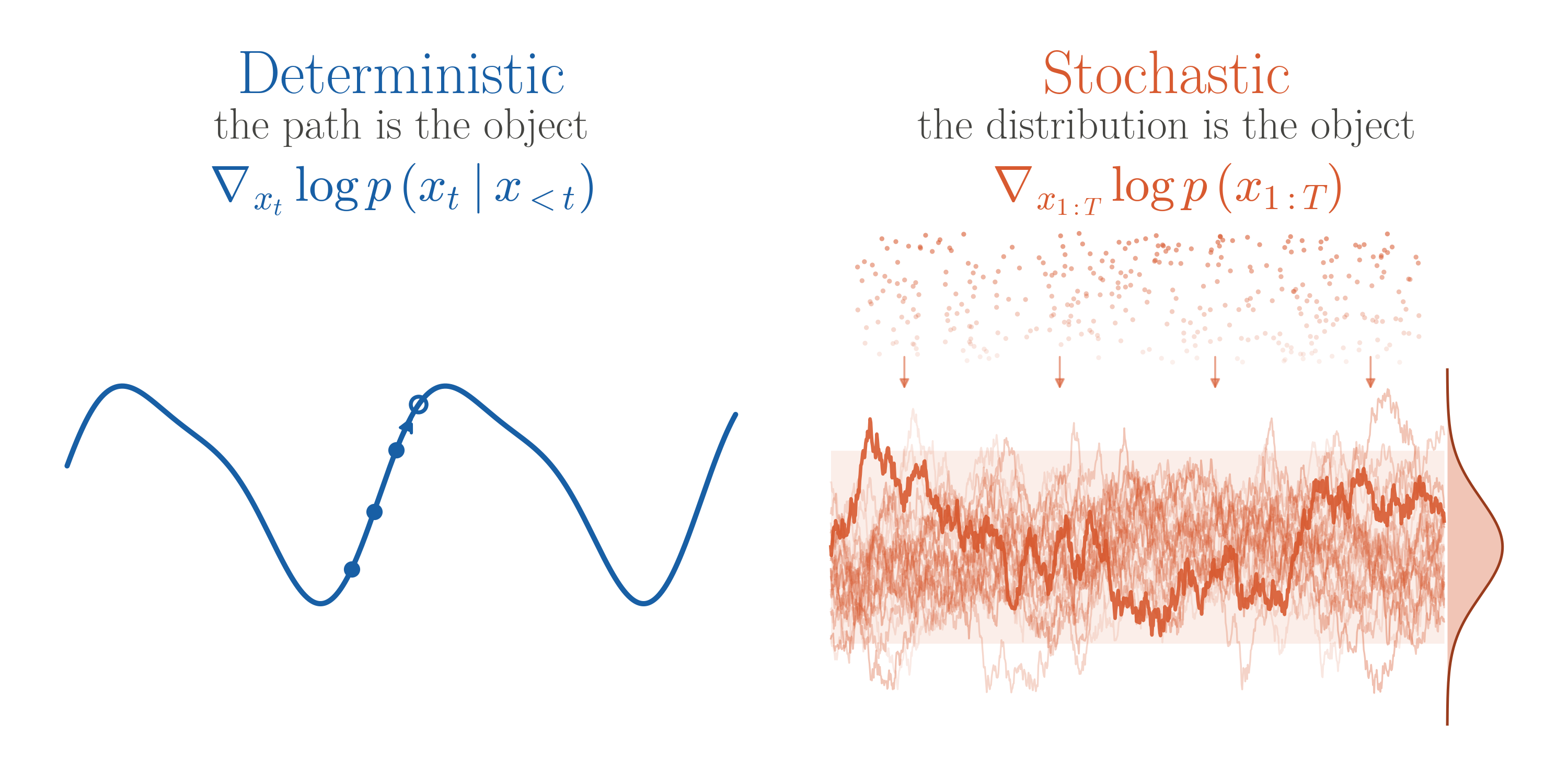}
\end{figure}

\section{Introduction}
\label{sec:intro}

The accurate characterization of dynamical processes in molecular systems depends
on the quality and quantity of available simulation data. In computational
chemistry, the relationship between statistical precision and sample size typically follows
the scaling law $\text{SE} = s/\sqrt{n}$, where the standard error (SE) is inversely
proportional to the square root of the number of independent samples $n$ and $s$
denotes the population standard deviation. This relationship imposes a stringent
requirement: a tenfold reduction in statistical uncertainty necessitates a
hundredfold increase in sampling, a demand that becomes prohibitive for systems of
realistic complexity, particularly when high-level electronic structure methods or
costly molecular dynamics (MD) simulations are required.

This statistical bottleneck motivates the development of generative models capable of synthesizing additional time-series data that preserve the underlying stochastic properties of the original trajectories. Unlike deterministic physical models that can be fitted to observed data, many processes of chemical interest, such as solvent reorganization dynamics, barrier-crossing events, and electronic energy gap fluctuations, exhibit inherently stochastic behavior that defies simple functional representation.\cite{hansen86,Nitzan2006} Being able to augment limited simulation datasets with synthetic trajectories that faithfully reproduce both short-time correlations and long-time decorrelation could therefore %
accelerate scientific discovery while maintaining statistical rigor.

Many experimentally accessible dynamical observables in molecular systems
can be cast in terms of
equilibrium two-time %
correlation functions in either the time-domain or frequency-domain.\cite{Green1954, Kubo1966, Gordon1968,berne70,Nitzan2006}
As a result, MD simulations are often carried out for the sole purpose of calculating such two-time correlation functions.
 Examples include the diffusion constant, which is given in terms of the %
 time integral of the velocity-velocity correlation function;\cite{Nitzan2006} the optical absorption spectrum, which is given in terms of the Fourier transform of the dipole-dipole correlation function;\cite{mukamel_book} the vibrational energy relaxation rate constant, which is given in terms of the Fourier transform of the force-force correlation function;\cite{Nitzan2006,egorov96b,navrotskaya07b,shi03c}
chemical reaction rate constants, which are given in terms of a time integral over the flux--flux correlation function\cite{yamamoto60,Miller1983,shi02}; and the spectral density function which is given in terms of the cosine Fourier transform the donor-acceptor energy gap autocorrelation function.\cite{walters16a,tong20,huang2026charge}

Importantly, equilibrium two-time correlation functions are measures of the statistics underlying the stochastic dynamics of molecular systems in thermal equilibrium. Specifically, given the correlation function $C_{AA} (t) = \langle A (t) A(0) \rangle_{eq} - \langle A \rangle_{eq}^2$, $C_{AA} (0)$ corresponds to the variance, $\sigma_{AA}^2 = \langle A^2\rangle_{eq} - \langle A \rangle_{eq}^2$,
and its lifetime is a measure of the correlation time, which sets the fluctuation time scale. As such, the calculation of such equilibrium two-time correlation functions requires averaging over a large number of equilibrium MD trajectories, and converges like $1/\sqrt{n}$, where $n$ is the number of independent dynamical segments averaged over.

In this paper we consider the question of whether or not it is possible to train a generative machine learning  (ML) model on a small sample of MD trajectories (one that would be too small for establishing a two-time correlation function at the desirable level of convergence), so that it can generate synthetic trajectories with the same statistics but at a lower cost, which can then be used to generate a better converged two-time correlation function without incurring the cost associated with generating additional MD trajectories.
To this end, we frame the analysis in terms of the following three questions:
\begin{enumerate}
\item To what extent can the statistics of equilibrium fluctuations be learned from a finite sample
and reproduced by synthetic trajectories?
\item Which class of generative model is best suited to the task?
\item By how much can such synthetic trajectories improve the estimate?
\end{enumerate}

In this context, the relationship of the present work to recent work by Markland and coworkers should be noted.\cite{Chen2020ML,Chen2023GFP,Kelly2025EqT,Hausman2026GMM} Specifically, those authors used ML models to accelerate the simulation
of linear and nonlinear electronic spectra in the condensed phase.
Specifically, the spectra were assumed to be given in terms of second-order
cumulant expansions of the optical response functions, which in turn are given in terms of the two-time correlation function of the transition frequency. ML models
were then used to generate a transition frequency for each given nuclear configuration along the MD trajectories, which was shown to reduce the number of excited-state calculations from tens of thousands to a few hundreds.~\cite{Chen2020ML,Chen2023GFP,Kelly2025EqT}
However, obtaining converged spectra still required averaging over a large number of trajectories. In another related study, Markland and coworkers used ML models to extract spectral densities (whose knowledge is equivalent to that of the transition frequency two-time correlation function) from {  experimental}  electronic two-dimensional spectra.\cite{Hausman2026GMM}
The current paper is distinctly different and complementary to those studies in that ML is used to lower the number of trajectories needed to obtain converged energy gap (analogous to the transition frequency) two-time correlation functions and the corresponding spectral densities. %

The remainder of this paper is organized as follows.
Equilibrium two-time correlation functions are defined in the time and frequency domains, and the conceptual framework of using generative models for time series prediction is introduced in Sec. \ref{sec:prelim}.
A detailed outline of the generative ML models under consideration in this paper is presented in Sec. \ref{sec:models}.
The two stochastic models used for testing the ability of the generative models under consideration to generate statistically reliable synthetic stochastic trajectories are presented in Sec. \ref{sec:testsys}.
The results of applying the generative models to the two test models are presented in Sec. \ref{sec:results} and discussed in Sec. \ref{sec:discussion}.
Finally, a summary of the main results and conclusions is given in Sec. \ref{sec:conclusions}.
Additional technical details, derivations and figures are provided in appendices and as supplementary information (SI).

\section{Preliminary considerations}
\label{sec:prelim}

\subsection{The two-time correlation function, spectral density and power spectral density}
\label{subsec:spectral}

For the sake of concreteness, we will focus on the commonly used protocol for generating a spectral density function from MD simulation data for a system that satisfies Gaussian statistics.\cite{huang2026charge,walters16a,tong20}

To this end, consider a molecular system undergoing charge transfer (CT), which corresponds to a transition between two diabatic electronic states, referred to as the donor (D) state and acceptor (A) state. The fact that the charge distribution in the donor state is different from that in the acceptor state implies that, generally speaking, for a given nuclear configuration, ${\bf R} \equiv (R_1,...,R_N)$, the donor potential energy, $V_D ({\bf R})$, differs from the acceptor potential energy, $V_A ({\bf R})$. Here, $R_j$ is the coordinate of the $j$-th nuclear degree of freedom (DOF), where $j=1,2,\ldots,N$.
The donor-acceptor energy gap in nuclear configuration ${\bf R}$ can then be defined as follows:
\begin{equation}
U({\bf R}) = V_A ({\bf R}) - V_D ({\bf R})
\label{eq:U}
\end{equation}

For a system in thermal equilibrium at the donor state, the nuclear configuration is time-dependent,
${\bf R}_t$, which makes the donor-acceptor energy gap time-dependent too, $U({\bf R}_t)$. Importantly, since initial nuclear positions and momenta are sampled from the Boltzmann distribution that corresponds to thermal equilibrium, the dynamics of $U({\bf R}_t)$ is stochastic rather than deterministic, where each choice of initial state gives rise to a different trajectory ${\bf R}_t$, and thereby a different realization of the time series traced by $U({\bf R}_t)$.\cite{chandler87,tuckerman10} Since the system is at equilibrium, $U({\bf R}_t)$ is also a  stationary stochastic process\cite{vankampen92,kubo83} where $U({\bf R}_t)$ fluctuates around its stationary equilibrium average value, $\langle U({\bf R}) \rangle$, where  $\langle \cdots \rangle$ corresponding to averaging over trajectories whose initial states where sampled based on a Boltzmann distribution at the donor state.

Assuming Gaussian statistics, such a stationary stochastic process can be fully defined by the corresponding average and two-time autocorrelation function:
\begin{equation}
C(t) = \langle U({\bf R}_t) U({\bf R}_0) \rangle - \langle U({\bf R}) \rangle^2
\label{eq:CUUt}
\end{equation}
The Gaussian nature of the statistics is often given in terms of a spectral density function, which in turn is given in terms of the cosine Fourier transform of $C(t)$:\cite{huang2026charge,walters16a,tong20}
\begin{equation}
J(\omega) = \beta \omega \int_0^\infty dt \, C_{UU}(t) \cos(\omega t)~~.
\label{eq:J-omega}
\end{equation}
Here, $\beta = 1/k_{B}T$ is the inverse temperature and $C_{UU}(t) \equiv C(t)$ of Eq.~\eqref{eq:CUUt}.
 Alternatively, the Gaussian statistics can also be expressed in terms of the power spectral density (PSD), which is defined by:\cite{tuckerman10}
\begin{equation}
I (\omega) = \lim_{T\to\infty} \frac{1}{2T} \big\langle |\tilde{U}_T(\omega)|^2 \big\rangle
~~,~~
\tilde{U}_T(\omega) = \int_{-T}^T dt e^{i \omega t} U(t)~.
\label{eq:PSD}
\end{equation}
It should be noted that there is a one-to-one correspondence between $C(t)$, $J(\omega)$ and $I(\omega)$. Thus, for a system that satisfies Gaussian statistics, either $C(t)$ or $J(\omega)$ or $I(\omega)$ completely defines the statistics of the system.

In practice, $C(t)$ is calculated from MD simulation data. For example, given a single MD trajectory of length $M \delta$, where $M$ is the number of time steps of length $\delta$ each, $C(t)$ can be estimated by:\cite{hansen86,tuckerman10}
\begin{eqnarray}
C(n\delta;M) &=& \frac{1}{M-n} \sum_{m=1}^{M-n} U({\bf R}_{m\delta})
U({\bf R}_{(m+n)\delta})
\nonumber \\
&&- \left( \frac{1}{M} \sum_{m=1}^M U({\bf R}_{m\delta}) \right)^2~~.
\label{eq:Ct_MD}
\end{eqnarray}
Eq. \eqref{eq:Ct_MD} is based on the ergodic hypothesis, according to which time averaging is equivalent to ensemble averaging. It should be noted that $C(n\delta;M)$ is an estimate of the actual $C(n \delta)$, which corresponds to the limit $M \rightarrow \infty$ of $C(n\delta;M)$:
\begin{equation}
C(n\delta) = \lim_{M \rightarrow \infty} C(n\delta;M)
\end{equation}
The rate at which $C(n\delta;M)$ converges to $C(n\delta)$ scales like $1/\sqrt{M}$, which typically requires relatively long MD trajectories for generating an estimate of $C(n \delta)$ with an error bar which is below an acceptable tolerance. Since numerical errors typically accumulate over time, averaging over multiple equilibrium trajectories that are long compared to the correlation time but sufficiently short so as to avoid extensive numerical error accumulation, is often preferable.\cite{tuckerman10} In such a case, the error bar is often estimated by arranging the multiple trajectories into blocks and calculating the standard deviation between the averages over different blocks.

\subsection{Generative model for time series prediction}

Contemporary ML approaches to deep learning are based on two principal architectures for generative modeling of sequential data:
\begin{enumerate}
\item {\em Autoregressive generative models}, which generate time-series elements
sequentially, conditioning each prediction on previous observations. This family encompasses convolutional neural networks (CNNs),
recurrent neural networks, and attention-based transformers \cite{Vaswani2017}.
\item {\em Simultaneous generative models}, which generate an entire sequence rather than predicting it one point at a time. Here we consider a variational autoencoder (VAE) model, a latent-variable generative model,\cite{Kingma2013} and a denoising diffusion probabilistic model (DDPM), which is a score-based generative model.\cite{Ho2020}
\end{enumerate}

In what follows, we evaluate the ability of two autoregressive generative models (CNN, Transformer) and two simultaneous generative models (VAE, DDPM) to accelerate the calculation of $C(t)$, $J(\omega)$ and $I (\omega)$ from MD simulation data.

\subsection{Using generative models for calculating $C(t)$, $J (\omega)$ and $I (\omega)$}

The standard protocol for calculating $C(t)$, $J (\omega)$ and $I (\omega)$ outlined in Sec. \ref{subsec:spectral} typically requires averaging over a large number of trajectories in order to bring the error bar to an acceptable level. The computational cost of doing so obviously grows with the number trajectories one has to average over for the error bar to reach an acceptable level. In this paper, we propose an alternative protocol, which is based on using a relatively small number of trajectories generated via MD simulations to train a generative model, which could then be used to generate additional synthetic stochastic time series with the same statistics, but at a much lower computational cost.

Specifically, %
let $N_r$ be the number of time series used to train the generative model and let $N_s$ be the number of synthetic time series generated by the generative model. Calculating $C(t)$, $J(\omega)$ and $I (\omega)$ based on the set of
$N_s$ synthetically generated time series, three outcomes are possible:
\begin{enumerate}
\item The error bar would remain the same with increasing $N_s$.
\item The error bar would decrease indefinitely with increasing $N_s$.

\item The error bar would decrease with increasing $N_s$ up to a critical value of $N_s$, and then remain the same or increase when $N_s$ is increased beyond that critical value.
\end{enumerate}

In what follows, we establish that for the generative models under consideration, outcome 3 is observed (see also Appendix~\ref{sec:gda_theory}). %

\section{Generative Models for Stochastic Trajectories}
\label{sec:models}

\subsection{Two-time correlation functions from generative models}
\label{subsec_C2}

We start by casting two-time correlation functions in terms of the probabilistic description underlying the generative model of choice.
To this end, consider a stationary stochastic process $\{ x(t) \}$, which stands for all the possible realizations of the stochastic process (as opposed to a deterministic process where there is only one such realization). Let the equilibrium probability density for the stochastic variable to have the value $x$, which for a stationary stochastic process is time-independent, be given by $\rho_{\mathrm{eq}} (x)$. Also let the conditional probability density for the stochastic variable to have the values $x_t$ a time $t$ provided that it has the value $x_0$ at time $0$ be given by $G(x_t,t\,|\,x_0,0)$. Then, the two-time correlation function in the time domain is given by:
\begin{equation}
  \begin{split}
    C (t)
    &=\; \int\! dx_0\!\int\! dx_t\;\, x_t\, x_0\; \rho_{\mathrm{eq}}(x_0)\, G(x_t,t\,|\,x_0,0)~~.
  \end{split}
  \label{eq:corr_cond}
\end{equation}
It should be noted that $p(x_0,x_t) = \rho_{\mathrm{eq}}(x_0)\, G(x_t,t\,|\,x_0,0)$ is the two-time joint probability density for $x=x_0$ at time $0$ {  and} $x=x_t$ at time t. At the same time, the generative model can be thought of as a representation of the $n$-time joint probability,
$p(x_0,x_{t_1},\ldots,x_{t_{n-1}})$, and as such can be used to generate higher than second order time correlation functions. However, in what follows we make the choice of choosing the two-time correlation function in Eq. \eqref{eq:corr_cond} as the learnable quantity.

\subsection{Sequential and simultaneous generation}
\label{sec:seqsim}

The four generative models considered and compared in this paper fall into two classes that differ with respect to how the aforementioned joint probability is handled.

{  Autoregressive models} synthesize data sequentially:
\begin{equation}
\hat{x}_{n} = f_\theta (x_0,x_1,\ldots,x_{n-1})~~.
\end{equation}
Here, $x_j \equiv x(t_j)$ is the value of the stochastic variable at the $j$-th point of a uniform time grid, $t_j = j\,\Delta t$, with $\Delta t$ the sampling interval,
$\hat{x}_{n}$ is the value predicted by the autoregressive model for $x_{n}$, given the known preceding values $(x_0,x_1,\ldots,x_{n-1})$,
and $f_\theta$ denotes the neural network realization of the model, where $\theta$ is the set of trainable parameters. It should be noted that in this case, one predicts the next step based on all the steps that preceded it, {  one step at a time}.

The sequential generation underlying autoregressive models can amplify errors when small errors per step can accumulate over multiple steps to yield divergent trajectories.\cite{Lin2025DiffBP}
In contrast, { simultaneous models} generate the entire sequence at once (as opposed to one step at a time):
\begin{equation}
    \hat{\mathbf{x}} = g_\theta(\mathbf{z}), \quad \mathbf{z} \sim p(\mathbf{z})~~,
    \label{eq:score}
\end{equation}
Here, $\mathbf{z}$ is a
random vector which is drawn from a fixed,
parameter-free, probability distribution $p(\mathbf{z})$ (a standard Gaussian for both simultaneous models considered here, see Sec.~\ref{sec:sb_models}), and $g_\theta$ denotes the neural network that maps $\mathbf{z}$ onto an entire sequence
$\hat{\mathbf{x}}= (\hat{x}_0,\ldots,\hat{x}_{n-1})$, where $\theta$ stands for the set of trainable parameters. Since $p(\mathbf{z})$ is fixed, everything the model learns about the data is stored in $\theta$, and each new draw of $\mathbf{z}$ yields a new synthetic sequence.
The global nature of the mapping in Eq. \eqref{eq:score} eliminates the aforementioned error accumulation to which autoregressive models are susceptible to and permits parallel computation during both training and inference.
As a result, simultaneous models can be expected to do better at capturing correlations across the entire sequence.

The distinction tested in this work is therefore between squared-error
sequential point prediction and whole-sequence generation.
The former learns a conditional mean, which under-disperses when the next value carries an irreducible stochastic innovation.
This does not by itself imply that probabilistic autoregressive models are unsuited to stochastic processes (the reader is referred to  Appendix~\ref{sec:arch_formal} for a more detailed discussion of this point).

\subsection{Autoregressive models}
\label{sec:ar_models}

Autoregressive models factorize the joint distribution of a trajectory of $L$ points,
$\mathbf{x} = (x_1, \ldots, x_L)$, into a product of conditional
distributions,
\begin{equation}
    p(\mathbf{x}) = \prod_{t=1}^{L} p(x_t \mid x_{t-1}, \ldots, x_1)~~,
\end{equation}
where $t$ is an integer time index and $p(x_t \mid x_{t-1}, \ldots, x_1)$ is the probability density of $x_t$ given all preceding values.
In practice the conditioning is truncated to a window of the $k$ most recent
values, and a neural network $f_\theta$ predicts the next value from that
window~\cite{Lim2021},
\begin{equation}
    \hat{x}_{t+1} = f_\theta(x_{t-k+1}, \ldots, x_t)~~,
\end{equation}
where the hat marks the model output, $\theta$ denotes the trainable
parameters, and $k$ is the length of the look back window ($k = 1024$ in this work, Sec.~\ref{sec:protocol}). The models used here output a single predicted
value and are trained on its squared error, so $f_\theta$ approximates the
conditional mean $\mathbb{E}[x_{t+1} \mid x_{t-k+1}, \ldots, x_t]$, where $\mathbb{E}[\,\cdot \mid \cdot\,]$ denotes a conditional expectation value.
Trajectories are then generated by feeding each prediction back as input.
\subsubsection{Convolutional neural network (CNN)}

Convolutional neural networks extract hierarchical features through localized filter operations. For one-dimensional time-series data, the convolution operation is:
\begin{equation}
    h_j^{(l)} = \sigma\left(\sum_{i=1}^{C_{l-1}} \mathbf{w}_{ij}^{(l)} * h_i^{(l-1)} + b_j^{(l)}\right)
\end{equation}
where $h_j^{(l)}$ denotes the $j$-th feature map (output channel) at layer $l$, a sequence along the time axis, with $h^{(0)}$ the input window (a single channel, $C_0=1$); $\mathbf{w}_{ij}^{(l)}$ represents the learnable convolutional kernel connecting input channel $i$ to output channel $j$ (here of width $5$ points); $*$ denotes the one-dimensional discrete convolution along the time axis; $b_j^{(l)}$ is the bias term; $\sigma(\cdot)$ is the activation function (here the rectified linear unit, $\sigma(y)=\max(0,y)$); and $C_{l-1}$ indicates the number of input channels from the previous layer. The model captures local temporal patterns through a hierarchy of increasingly abstract representations. In the model used here, each of three convolutional layers is followed by max pooling, which halves the length of the feature maps, and fully connected layers map the final feature maps onto the single predicted value $\hat{x}_{t+1}$ (Supporting Information, Table~S1).

\subsubsection{Attention-based transformer}

The transformer architecture eschews recurrence in favor of self-attention mechanisms that directly model interactions between all time points. The scaled dot-product attention is computed as:
\begin{equation}
    \text{Attention}(\mathbf{Q}, \mathbf{K}, \mathbf{V}) = \text{softmax}\left(\frac{\mathbf{Q}\mathbf{K}^\top}{\sqrt{d_k}}\right)\mathbf{V}
\end{equation}
Here, each of the $k$ values of the input window is first mapped by a learned linear transformation onto a $d_{\text{model}}$-dimensional vector, its embedding, and the $k$ embeddings form the rows of a $k \times d_{\text{model}}$ matrix $\mathbf{X}$.
$\mathbf{Q}$, $\mathbf{K}$, and $\mathbf{V}$ represent the query, key, and value matrices, whose $m$-th rows are the query, key, and value vectors of the $m$-th time point of the window; they are derived from $\mathbf{X}$ through learned linear projections (see below). $d_k$ is the dimensionality of the query and key vectors, $^\top$ denotes the matrix transpose, and the softmax function is applied to each row of its argument, $\text{softmax}(\mathbf{a})_m = e^{a_m}/\sum_{m'} e^{a_{m'}}$, so that each row of the output is a weighted average of the value vectors, with weights set by the similarity between the corresponding query and the keys. Multi-head attention extends this by computing attention in parallel across multiple subspaces:
\begin{multline}
    \text{MultiHead}(\mathbf{Q}, \mathbf{K}, \mathbf{V}) \\
    = \text{Concat}(\text{head}_1, \ldots, \text{head}_h)\mathbf{W}^O
\end{multline}
with each head defined as
\begin{equation}
    \text{head}_i = \text{Attention}(\mathbf{Q}\mathbf{W}_i^Q, \mathbf{K}\mathbf{W}_i^K, \mathbf{V}\mathbf{W}_i^V).
\end{equation}
Here, $h$ is the number of attention heads, $i = 1,\ldots,h$;
$\mathbf{W}_i^Q, \mathbf{W}_i^K \in \mathbb{R}^{d_{\text{model}} \times d_k}$ and $\mathbf{W}_i^V \in \mathbb{R}^{d_{\text{model}} \times d_v}$ are the learned projection matrices of head $i$, with $d_v$ the dimensionality of the value vectors;
$\text{Concat}$ places the $h$ head outputs side by side to form a $k \times h d_v$ matrix;
and $\mathbf{W}^O \in \mathbb{R}^{h d_v \times d_{\text{model}}}$ is a learned output projection matrix that maps the concatenated heads back onto the $d_{\text{model}}$-dimensional embedding space.
In self-attention, as used here, $\mathbf{Q} = \mathbf{K} = \mathbf{V} = \mathbf{X}$. Because the model must predict the future from the past, a causal mask is applied, setting the entries of $\mathbf{Q}\mathbf{K}^\top$ above the diagonal to $-\infty$ before the softmax, so that each time point attends only to itself and to earlier time points.
Positional encodings are added to the input embeddings to convey temporal ordering:
\begin{equation}
    \text{PE}_{(pos, 2i)} = \sin(pos/10000^{2i/d_{\text{model}}})
\end{equation}
\begin{equation}
    \text{PE}_{(pos, 2i+1)} = \cos(pos/10000^{2i/d_{\text{model}}})
\end{equation}
where $pos = 0,\ldots,k-1$ denotes the temporal position within the window and $i = 0,\ldots,d_{\text{model}}/2-1$ indexes pairs of embedding dimensions.
In the model used here, $d_{\text{model}} = 128$, $h = 4$, and $d_k = d_v = d_{\text{model}}/h = 32$. Two such attention layers are stacked, each followed by a position-wise feed-forward network and each wrapped in a residual connection and layer normalization, and the predicted value $\hat{x}_{t+1}$ is read out from the output at the last position of the window (Supporting Information, Table~S2).

\subsection{Simultaneous generative models}
\label{sec:sb_models}

The two simultaneous models generate a whole trajectory window rather than predicting one point at a time. The VAE model does so through a latent variable representation, while the DDPM is a score-based model, which reframes generation as learning the gradient of the log-probability density, the score function, and generates by following that gradient field.

\subsubsection{Variational autoencoder (VAE)}

Variational autoencoders \cite{Kingma2013} learn a compressed latent representation of the data through an encoder-decoder architecture with a probabilistic regularization. For a training window $\mathbf{x}$, the loss minimized during training is
\begin{equation}
    \mathcal{L}_{\text{VAE}} = -\mathbb{E}_{q_\phi(\mathbf{z}|\mathbf{x})}[\log p_\theta(\mathbf{x}|\mathbf{z})] + \beta_{\mathrm{KL}} D_{\text{KL}}(q_\phi(\mathbf{z}|\mathbf{x}) \,\|\, p(\mathbf{z}))~~.
\end{equation}
Here, $\mathbf{z}$ is a latent vector of much lower dimension than $\mathbf{x}$ ($32$ here, compared with $1024$ points); the encoder $q_\phi(\mathbf{z}|\mathbf{x})$, with trainable parameters $\phi$, is the approximate posterior, i.e., the probability density of $\mathbf{z}$ given $\mathbf{x}$; the decoder $p_\theta(\mathbf{x}|\mathbf{z})$, with trainable parameters $\theta$, is the probability density of $\mathbf{x}$ given $\mathbf{z}$; and $\mathbb{E}_{q_\phi(\mathbf{z}|\mathbf{x})}[\cdots]$ denotes an average over $\mathbf{z}$ drawn from $q_\phi(\mathbf{z}|\mathbf{x})$.
The first term represents the reconstruction loss, which for the Gaussian decoder used here reduces, up to constants, to the squared error between $\mathbf{x}$ and its reconstruction. The second term is the Kullback--Leibler divergence, $D_{\text{KL}}(q\,\|\,p) = \int d\mathbf{z}\, q(\mathbf{z}) \ln[q(\mathbf{z})/p(\mathbf{z})]$, between the approximate posterior and the prior $p(\mathbf{z})$, chosen here as a standard Gaussian $\mathcal{N}(0, \mathbf{I})$ with $\mathbf{I}$ the identity matrix, and $\beta_{\mathrm{KL}}$ is its weight ($\beta_{\mathrm{KL}} = 10^{-3}$ in this work; for $\beta_{\mathrm{KL}} = 1$, $-\mathcal{L}_{\text{VAE}}$ is the evidence lower bound on $\log p_\theta(\mathbf{x})$). The reparameterization trick enables differentiable sampling:
\begin{equation}
    \mathbf{z} = \boldsymbol{\mu}_\phi(\mathbf{x}) + \boldsymbol{\sigma}_\phi(\mathbf{x}) \odot \boldsymbol{\epsilon}, \quad \boldsymbol{\epsilon} \sim \mathcal{N}(0, \mathbf{I})
\end{equation}
where $\boldsymbol{\mu}_\phi$ and $\boldsymbol{\sigma}_\phi$ are the mean and standard deviation of $q_\phi(\mathbf{z}|\mathbf{x})$ output by the encoder network, and $\odot$ denotes element-wise multiplication. Once trained, the encoder is discarded, and synthetic windows are generated by drawing $\mathbf{z} \sim \mathcal{N}(0, \mathbf{I})$ and passing it through the decoder network, which plays the role of $g_\theta$ in Eq.~\eqref{eq:score}.

\subsubsection{Denoising diffusion probabilistic model (DDPM)}

Denoising diffusion probabilistic models \cite{Ho2020} define a forward diffusion process that gradually corrupts data with Gaussian noise over $N_{\mathrm{d}}$ noising steps, $j = 1,\ldots,N_{\mathrm{d}}$:
\begin{equation}
    q(\mathbf{x}_j | \mathbf{x}_{j-1}) = \mathcal{N}(\mathbf{x}_j; \sqrt{1 - \beta_j}\mathbf{x}_{j-1}, \beta_j\mathbf{I})
\end{equation}
Here, $\mathbf{x}_0$ is a clean training window and $\mathbf{x}_j$ is its noised version after $j$ steps (the index $j$ labels the noising step, not physical time); $\mathcal{N}(\mathbf{y}; \boldsymbol{\mu}, \boldsymbol{\Sigma})$ denotes the Gaussian probability density of $\mathbf{y}$ with mean $\boldsymbol{\mu}$ and covariance matrix $\boldsymbol{\Sigma}$; $\mathbf{I}$ is the identity matrix; and $\beta_j \in (0,1)$ follows a predefined noise schedule. Here $N_{\mathrm{d}} = 1000$ and $\beta_j$ increases linearly from $10^{-4}$ to $0.02$, so that $\mathbf{x}_{N_{\mathrm{d}}}$ is essentially pure Gaussian noise. The reverse generative process learns to denoise:
\begin{equation}
    p_\theta(\mathbf{x}_{j-1} | \mathbf{x}_j) = \mathcal{N}(\mathbf{x}_{j-1}; \boldsymbol{\mu}_\theta(\mathbf{x}_j, j), \boldsymbol{\Sigma}_\theta(\mathbf{x}_j, j))
\end{equation}
Training minimizes a simplified objective:
\begin{equation}
    \mathcal{L}_{\text{simple}} = \mathbb{E}_{j, \mathbf{x}_0, \boldsymbol{\epsilon}}\left[||\boldsymbol{\epsilon} - \boldsymbol{\epsilon}_\theta(\mathbf{x}_j, j)||^2_2\right]
\end{equation}
where the average is over $j$ drawn uniformly from $1,\ldots,N_{\mathrm{d}}$, over training windows $\mathbf{x}_0$, and over $\boldsymbol{\epsilon} \sim \mathcal{N}(0, \mathbf{I})$, the noise added at step $j$; $\boldsymbol{\epsilon}_\theta$ is the noise predicted by the neural network (here a one-dimensional U-Net, Supporting Information, Table~S4); and $\|\cdot\|_2$ is the Euclidean norm. Using the reparameterization $\mathbf{x}_j = \sqrt{\bar{\alpha}_j}\mathbf{x}_0 + \sqrt{1 - \bar{\alpha}_j}\boldsymbol{\epsilon}$ with $\bar{\alpha}_j = \prod_{s=1}^j (1 - \beta_s)$, the model learns to predict and remove the noise component.
The predicted noise determines the mean of the reverse step, $\boldsymbol{\mu}_\theta(\mathbf{x}_j, j) = [\mathbf{x}_j - \beta_j\, \boldsymbol{\epsilon}_\theta(\mathbf{x}_j, j)/\sqrt{1-\bar{\alpha}_j}\,]/\sqrt{1-\beta_j}$, while the covariance is held fixed, $\boldsymbol{\Sigma}_\theta = \beta_j \mathbf{I}$.
The predicted noise is also, up to a factor, an estimate of the score, $\nabla_{\mathbf{x}_j} \log q(\mathbf{x}_j) \approx -\boldsymbol{\epsilon}_\theta(\mathbf{x}_j, j)/\sqrt{1-\bar{\alpha}_j}$, which is why the DDPM is a score-based model.
A synthetic window is generated by drawing $\mathbf{x}_{N_{\mathrm{d}}} \sim \mathcal{N}(0, \mathbf{I})$, which plays the role of $\mathbf{z}$ in Eq.~\eqref{eq:score}, and applying the reverse step $N_{\mathrm{d}}$ times, down to $\mathbf{x}_0$.

\section{Stochastic Test Models}
\label{sec:testsys}

In this section, we outline the two stochastic test models that we will use to test the ability of the four generative models outlined in Sec. \ref{sec:models} to generate synthetic time series with the same statistics as the training set, as measured by $C(t)$, $J(\omega)$ and $I(\omega)$.

\subsection{The Ornstein--Uhlenbeck (OU) process}
\label{sec:sys_ou}

The Ornstein--Uhlenbeck (OU) process\cite{uhlenbeck30} is a Gaussian stationary Markov process
with an exponential two-time correlation function. The dynamics of the OU
stochastic variable $\xi(t)$ is governed by the following equation of motion:
\begin{equation}
    \dot{\xi}(t) = -\frac{1}{\tau}\,\xi(t) + \frac{\sqrt{2D}}{\tau}\,\Gamma(t)~~.
\end{equation}
Here, $\tau$ is the correlation time, $D$ sets the stochastic noise strength, and
$\Gamma(t)$ is Gaussian white noise such that
$\langle\Gamma(t)\Gamma(t')\rangle = \delta(t-t')$. The stationary variance is
$\langle\xi^2\rangle = D/\tau$, and the normalized autocorrelation function is
\begin{equation}
    C(t) = \frac{\langle\xi(t)\xi(0)\rangle}{\langle\xi^2\rangle} = e^{-|t|/\tau}~~.
\end{equation}

In what follows, stochastic OU time series were generated via the exact integral algorithm of Fox
et al.~\cite{Fox1988}. Within this algorithm:
\begin{equation}
\xi_{n+1} = \xi_n e^{-\Delta t/\tau} + h_n~~,
\label{eq:OU_alg}
\end{equation}
where $\Delta t$ is the time step and $h_n$ is a random variable sampled from a Gaussian probability density with
variance $(D/\tau)(1-e^{-2\Delta t /\tau})$.
It should be noted that the update is exact for any step size.

A stochastic time series consisting of $10^{5}$ time steps was generated by the aforementioned algorithm, where we set $\Delta t = 10^{-3}$
and $D = 1$, and
varied $\tau$ from $10^{-3}$ to $1$, so that
$\tau/\Delta t = 1-10^{3}$. %
The training set was set to a $1,024$-point window.
Setting the variance to unity, left $\tau/\Delta t$ as the only parameter.
For this model, times and frequencies are reported in units of $\Delta t$ and $1/\Delta t$, respectively.

\subsection{Electron-transfer energy gap (ETG) process}
\label{sec:sys_etg}

The second stochastic process we will employ %
is based on the CT reaction scenario outlined in Sec. \ref{subsec:spectral}. To this end, we will train the generative models on trajectories generated by all-atom MD simulations of a peptoid chain consisting of nine N-(S)-(1-phenylethyl)glycine (nspe) units and functionalized with Fe$^{2+}$-porphyrin and Fe$^{3+}$-porphyrin complexes in positions 4 and 7, and dissolved in liquid acetonitrile, at room temperature.
\begin{equation}
  \ce{Fe^{II}-P + Fe^{III}-P -> Fe^{III}-P + Fe^{II}-P},
  \label{eq:ct}
\end{equation}

The CT reaction in this case occurs between the Fe$^{2+}$-porphyrin and Fe$^{3+}$-porphyrin complexes, and the stochastic variable corresponds to the donor-acceptor energy gap, $U(t)$ (see Eq. \eqref{eq:U}).
In this case, we will test the robustness of the generative models
on fifteen different peptoid conformations.

\section{Results}
\label{sec:results}

The following four generative models outlined in Sec. \ref{sec:models} were evaluated on the two test stochasic processes (OU and ETG) described in Sec. \ref{sec:testsys}:
\begin{enumerate}
\item The Denoising Diffusion Probabilistic Model (DDPM).
\item The Variational Autoencoder (VAE).
\item The Convolutional Neural Network (CNN).
\item The decoder-only Transformer (Transformer).
\end{enumerate}

Stochastic variables were normalized to have zero mean and unit variance prior to training.
The performance of the generative models was estimated based on their ability to generate synthetic trajectories that can reproduce the two-time autocorrelation function,
$C(t)$ (see Eq. \eqref{eq:CUUt}),
the spectral density $J(\omega)$ (See Eq. \eqref{eq:J-omega}),
and the PSD $I (\omega)$ (see Eq. \eqref{eq:PSD}).

\subsection{Protocol}
\label{sec:protocol}

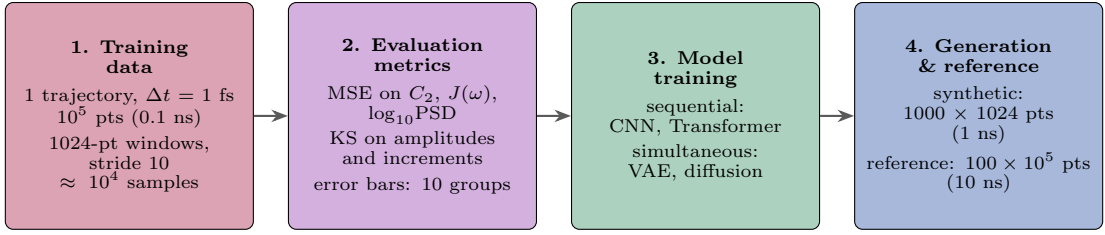
\begin{figure*}[ht]
  \centering
  \definecolor{stageA}{HTML}{DBA3B4}
  \definecolor{stageB}{HTML}{D3B0DE}
  \definecolor{stageC}{HTML}{ACD1BE}
  \definecolor{stageD}{HTML}{A7B8DB}
  \begin{tikzpicture}[
      node distance=0.45cm,
      stage/.style={draw, rounded corners=3pt, fill=#1, align=center,
                    text width=3.0cm, minimum height=3.0cm, inner sep=4pt,
                    font=\footnotesize, execute at begin node={\hyphenpenalty=10000}},
      arr/.style={-{Stealth[length=2.2mm]}, thick, gray!70!black}]
    \node[stage=stageA] (data) {{\bfseries 1. Training}\\{\bfseries data}\\[3pt]
      1 trajectory, $\Delta t = 1$~fs \\
      $10^{5}$ pts (0.1~ns) \\[2pt]
      1024-pt windows,\\ stride 10\\ $\approx10^{4}$ samples};
    \node[stage=stageB, right=of data] (metric) {{\bfseries 2. Evaluation}\\{\bfseries metrics}\\[3pt]
      MSE on $C_2$, $J(\omega)$,\\ $\log_{10}$PSD\\[2pt]
      KS on amplitudes\\ and increments\\[2pt]
      error bars: 10 groups};
    \node[stage=stageC, right=of metric] (train) {{\bfseries 3. Model}\\{\bfseries training}\\[3pt]
      sequential:\\ CNN, Transformer\\[2pt]
      simultaneous:\\ VAE, diffusion};
    \node[stage=stageD, right=of train] (gen) {{\bfseries 4. Generation}\\{\bfseries \& reference}\\[3pt]
      synthetic: $1000 \times 1024$ pts\\ (1~ns)\\[2pt]
      reference: $100 \times 10^{5}$ pts\\ (10~ns)};
    \draw[arr] (data) -- (metric);
    \draw[arr] (metric) -- (train);
    \draw[arr] (train) -- (gen);
  \end{tikzpicture}
  \caption{Protocol shared by all models and both data sources
  (Section~\ref{sec:protocol}).}
  \label{fig:protocol}
\end{figure*}

All results are obtained with a single protocol, applied identically to the
four models and to both data sources (Fig.~\ref{fig:protocol}).

\paragraph{Training data.}
Each model is trained on a single trajectory of $L=10^{5}$ points, for both
the OU process and ETG; for ETG the sampling interval is $\Delta t = 1$~fs
($0.1$~ns). The trajectory
is normalized to zero mean and unit variance and cut into windows of $W=1024$
points with a stride $s$, which gives
\begin{equation}
  N_w=\left\lfloor\frac{L-W}{s}\right\rfloor+1
  \label{eq:n_windows}
\end{equation}
training samples, $N_w\approx10^{4}$ at the default $s=10$. For the
autoregressive models each window is extended by one point, the prediction
target. The choice of stride is examined in Section~\ref{sec:diffsweep}.

\paragraph{Evaluation metrics.}
Each ensemble is summarized by three curve statistics --- $C_2(\tau)$,
$J(\omega)$, and the PSD --- and by the distributions
of values and of one-step increments. A curve statistic $f$ on a grid
$\{s_m\}_{m=1}^{M}$ of lags or frequencies is scored by
\begin{equation}
  \mathrm{MSE}_f = \frac{1}{M}\sum_{m=1}^{M}
  \big[\bar f_{\mathrm{gen}}(s_m)-\bar f_{\mathrm{real}}(s_m)\big]^{2},
  \label{eq:mse_metric}
\end{equation}
where the bar denotes the ensemble average and the PSD enters as
$\log_{10}\mathrm{PSD}$. Distributions are compared by the Kolmogorov--Smirnov
statistic
\begin{equation}
  \mathrm{KS} = \sup_{y}\big|F_{\mathrm{gen}}(y)-F_{\mathrm{real}}(y)\big|,
  \label{eq:ks_metric}
\end{equation}
with $F$ the empirical cumulative distribution of the amplitudes
(KS$_{\mathrm{amp}}$) or increments (KS$_{\mathrm{inc}}$); it ranges from $0$
(identical) to $1$ (disjoint). Lower is better for every metric. Error bars
on the curves are the standard deviation of a statistic estimated from about
$10^{5}$ points, the size of the training set ($0.1$~ns for ETG): for a
generated ensemble, over ten groups of $100$ trajectories; for the reference,
over its $100$ individual trajectories. Where not visible, they are smaller
than the line width. These error bars describe the statistical uncertainty
of the curves within a single generated ensemble. The scalar metrics also
vary between independent generation draws from the same trained model
(Section~\ref{sec:diffsweep}), so every entry in
Tables~\ref{tab:crossmodel_expnoise}--\ref{tab:structures} is the median
over five such draws. These metrics are used only to evaluate trained
models and are not the objectives the models are trained on, which are given
below.

\paragraph{Model training.}
The sequential models, a one-dimensional CNN and a Transformer, map a
$k$-point window to the next point (Section~\ref{sec:ar_models}) and are
trained on the squared error of that prediction,
\begin{equation}
  \mathcal{L}_{\mathrm{AR}}(\theta) =
  \big\langle \big[f_\theta(x_{t-k+1}, \ldots, x_t) - x_{t+1}\big]^2 \big\rangle ;
  \label{eq:loss_ar}
\end{equation}
they generate by feeding their own predictions back as input, starting from a random window from testing set. The simultaneous models learn the joint distribution
of a whole window $\mathbf{x}$. The VAE, with encoder
$q_\phi(\mathbf{z}\mid\mathbf{x})$ and decoder $p_\theta(\mathbf{x}\mid\mathbf{z})$,
is trained by maximizing the evidence lower bound, that is, by minimizing
\begin{equation}
  \begin{split}
  \mathcal{L}_{\mathrm{VAE}}(\theta,\phi) = {}&
  \Big\langle -\mathbb{E}_{q_\phi(\mathbf{z}\mid\mathbf{x})}
  \big[\log p_\theta(\mathbf{x}\mid\mathbf{z})\big] \\
  &+ \beta_{\mathrm{KL}}\, D_{\mathrm{KL}}\big(q_\phi(\mathbf{z}\mid\mathbf{x})\,\|\,p(\mathbf{z})\big)
  \Big\rangle ,
  \end{split}
  \label{eq:loss_vae}
\end{equation}
with $\beta_{\mathrm{KL}} = 10^{-3}$ (Section~\ref{sec:sb_models}), and generates by decoding latent vectors drawn from the standard normal prior
$p(\mathbf{z})$. The diffusion model is a denoising diffusion probabilistic
model in which a one-dimensional U-Net $\boldsymbol{\epsilon}_\theta$ is
trained to recover the noise added to a window at step $j$ of a Gaussian
noising process,
\begin{equation}
  \mathcal{L}_{\mathrm{DDPM}}(\theta) =
  \Big\langle \big\| \boldsymbol{\epsilon} - \boldsymbol{\epsilon}_\theta
  \big(\sqrt{\bar\alpha_j}\,\mathbf{x} + \sqrt{1-\bar\alpha_j}\,\boldsymbol{\epsilon},\, j\big)
  \big\|^2 \Big\rangle ,
  \label{eq:loss_ddpm}
\end{equation}
with $\boldsymbol{\epsilon} \sim \mathcal{N}(0, I)$, $j$ drawn uniformly from
the $N_{\mathrm{d}}$ noising steps, and $\bar\alpha_j$ the cumulative noise
schedule; it generates by running the reverse process from Gaussian noise. In
each case $\langle\cdot\rangle$ averages over training windows. None of these
objectives involves the correlation function or the spectrum, so agreement in
those statistics tests what the models have learned rather than what they were
fitted to. Architectures, noise schedules, and training details are given in
the Supporting Information.

\paragraph{Generation and reference.}
Each model generates $1000$ synthetic trajectories of $1024$ points, a length
set by the architecture of the simultaneous models and by the instability of
long autoregressive rollouts. The reference is computed from $100$
independent real trajectories ($10$~ns), one hundred times the training data,
evaluated on $1024$-point windows so that real and synthetic statistics share
the same grids.
The data are partitioned as follows. The training trajectory is one of the
$100$ reference trajectories, so it contributes $1\%$ of the reference data.
The autoregressive models start each rollout from a $1024$-point window drawn
at random from a second trajectory, which is not used for training (it is
also one of the reference trajectories); rollouts
whose matched true continuation would extend beyond the end of that
trajectory are discarded (about $1\%$), so these models contribute about
$990$ rather than $1000$ trajectories. The simultaneous models generate
unconditionally and need no input at generation, so no separate test set is
required for them. The real-data curve of Section~\ref{sec:convergence} is
computed from a further $96$ trajectories that are disjoint from both the
training and the reference data. Rarely, a single diffusion draw diverges
during sampling, a numerical instability of the batch-normalization layers of
the U-Net; such draws are discarded and redrawn (no run used here produced
more than $5$ in $1000$; Supporting Information, Section~3.1).

\subsection{Cross-model comparison}
\label{sec:crossmodel}

This subsection addresses the first two questions of the Introduction:
whether the statistics of equilibrium fluctuations can be learned from a
sample too small to converge them, and which class of generative model is
suited to the task. Each model sees one hundredth of the data that defines
the reference (Section~\ref{sec:protocol}).

\subsubsection{Ornstein--Uhlenbeck (OU) process}

For a variance-normalized OU process, the correlation time sets the one-step
correlation $\rho=e^{-\Delta t/\tau}$ and hence how predictable the signal
is: nearly deterministic for $\tau\gg\Delta t$, nearly white noise for
$\tau\lesssim\Delta t$. Sweeping $\tau$ thus varies exactly the property that
separates sequential from simultaneous generation
(Section~\ref{sec:arch_formal}).

\begin{figure*}[t]
  \centering
  \includegraphics[width=\linewidth]{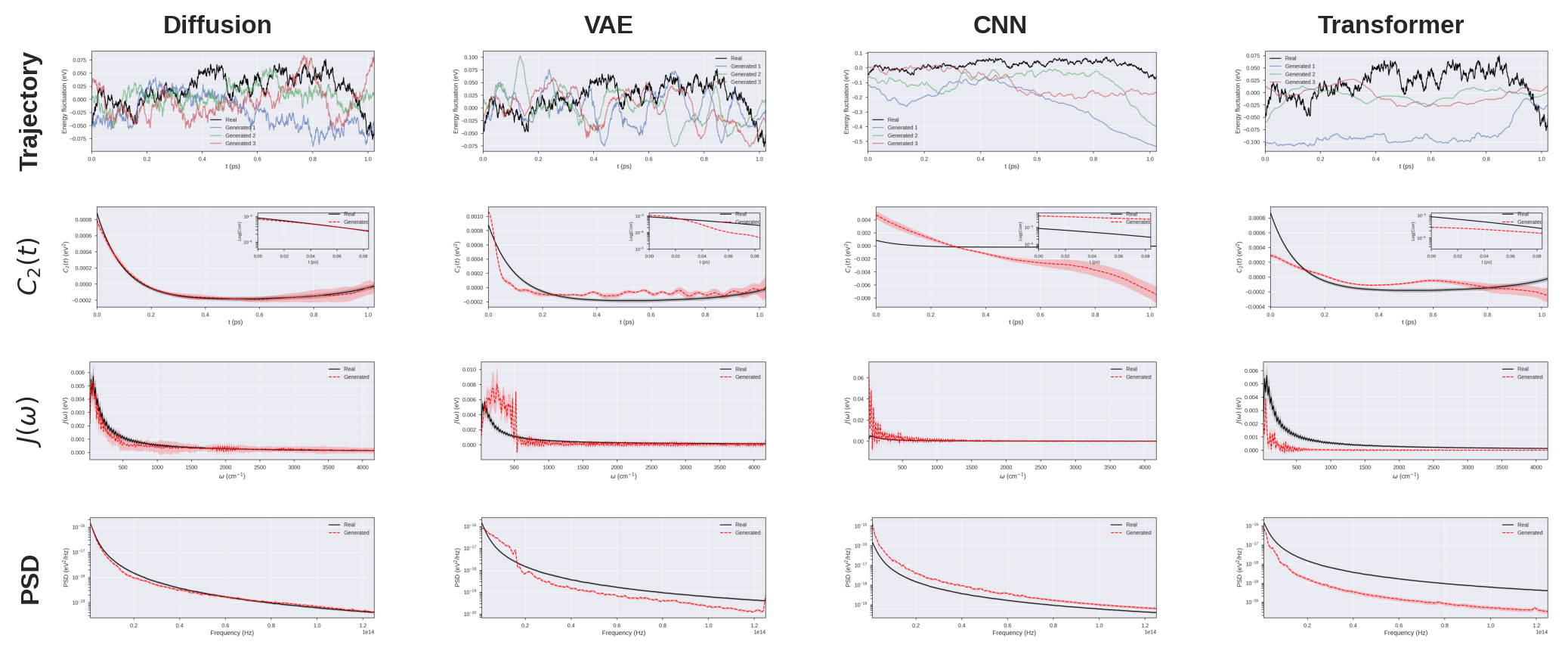}
  \caption{Generated statistics for the OU process ($D=1.0$,
  $\tau=10^{-1}$). Columns: models. Rows: a representative trajectory,
  $C_2(\tau)$, $J(\omega)$, and the PSD. Black: real reference (100
  trajectories); red: generated (1000 trajectories, a single generation draw).
  Error bars as defined in Section~\ref{sec:protocol}.}
  \label{fig:data_expnoise}
\end{figure*}

\begin{figure*}[t]
  \centering
  \includegraphics[width=\linewidth]{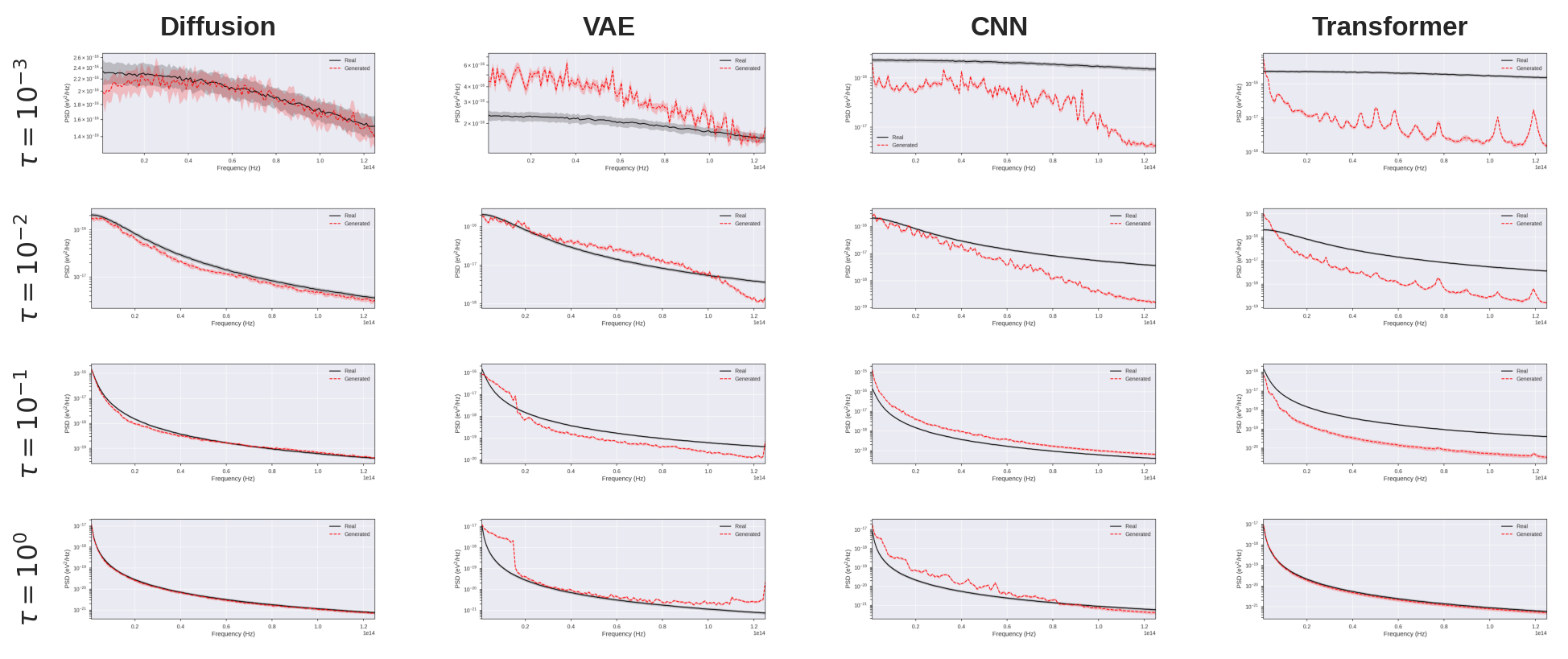}
  \caption{Generated PSD for the OU process across correlation times (rows,
  $\tau=10^{-3}$ to $10^{0}$) for the four models (columns). Black: real;
  red: generated.}
  \label{fig:tau_psd}
\end{figure*}

The results follow this distinction (Table~\ref{tab:crossmodel_expnoise},
Figs.~\ref{fig:data_expnoise} and~\ref{fig:tau_psd}). The diffusion model
reproduces the spectrum at every correlation time, with an error essentially
independent of $\tau$. The autoregressive models are competitive only at long
$\tau$; as $\tau\to0$ their spectrum collapses, because a conditional-mean predictor cannot anticipate unstructured noise and
returns the conditional mean. The VAE reproduces the amplitude distribution
but not the spectrum.

\begin{table*}[t]
  \centering
  \caption{OU process ($D=1.0$): KS$_{\mathrm{amp}}$ and MSE\,$\log_{10}$PSD
  (Eqs.~\eqref{eq:mse_metric}--\eqref{eq:ks_metric}) versus $\tau$. Each
  entry is the median of five generation draws from the same trained model,
  with only the sampling seed varied (seeds as in Table~\ref{tab:diffsweep}).
  Best per column in bold.}
  \label{tab:crossmodel_expnoise}
  \begin{tabular}{lrrrr}
\toprule
\multicolumn{5}{c}{\textbf{KS amplitude}} \\
\midrule
\textbf{Method} & \textbf{$\tau{=}10^{-3}$} & \textbf{$\tau{=}10^{-2}$} & \textbf{$\tau{=}10^{-1}$} & \textbf{$\tau{=}10^{0}$} \\
\midrule
\textbf{Diffusion} & $4.70\times10^{-3}$ & $1.99\times10^{-2}$ & $2.42\times10^{-2}$ & $3.13\times10^{-2}$ \\
\textbf{VAE} & {\boldmath$4.00\times10^{-3}$} & {\boldmath$2.59\times10^{-3}$} & {\boldmath$1.27\times10^{-2}$} & {\boldmath$1.04\times10^{-2}$} \\
\textbf{CNN} & $2.38\times10^{-1}$ & $9.43\times10^{-2}$ & $2.67\times10^{-1}$ & $1.27\times10^{-1}$ \\
\textbf{Transformer} & $3.72\times10^{-1}$ & $1.04\times10^{-1}$ & $2.20\times10^{-1}$ & $2.69\times10^{-1}$ \\
\bottomrule
\end{tabular}

\vspace{0.6em}

\begin{tabular}{lrrrr}
\toprule
\multicolumn{5}{c}{\textbf{MSE logPSD}} \\
\midrule
\textbf{Method} & \textbf{$\tau{=}10^{-3}$} & \textbf{$\tau{=}10^{-2}$} & \textbf{$\tau{=}10^{-1}$} & \textbf{$\tau{=}10^{0}$} \\
\midrule
\textbf{Diffusion} & {\boldmath$4.55\times10^{-4}$} & {\boldmath$6.89\times10^{-3}$} & {\boldmath$1.81\times10^{-3}$} & {\boldmath$3.28\times10^{-3}$} \\
\textbf{VAE} & $3.90\times10^{-1}$ & $2.90\times10^{-1}$ & $1.72\times10^{-1}$ & $5.29\times10^{-1}$ \\
\textbf{CNN} & $1.07\times10^{0}$ & $1.45\times10^{0}$ & $6.22\times10^{-2}$ & $6.27\times10^{-2}$ \\
\textbf{Transformer} & $6.06\times10^{0}$ & $2.25\times10^{0}$ & $1.22\times10^{0}$ & $8.74\times10^{-3}$ \\
\bottomrule
\end{tabular}

\end{table*}

\subsubsection{Energy-gap (ETG) trajectories}

\begin{figure*}[t]
  \centering
  \includegraphics[width=\linewidth]{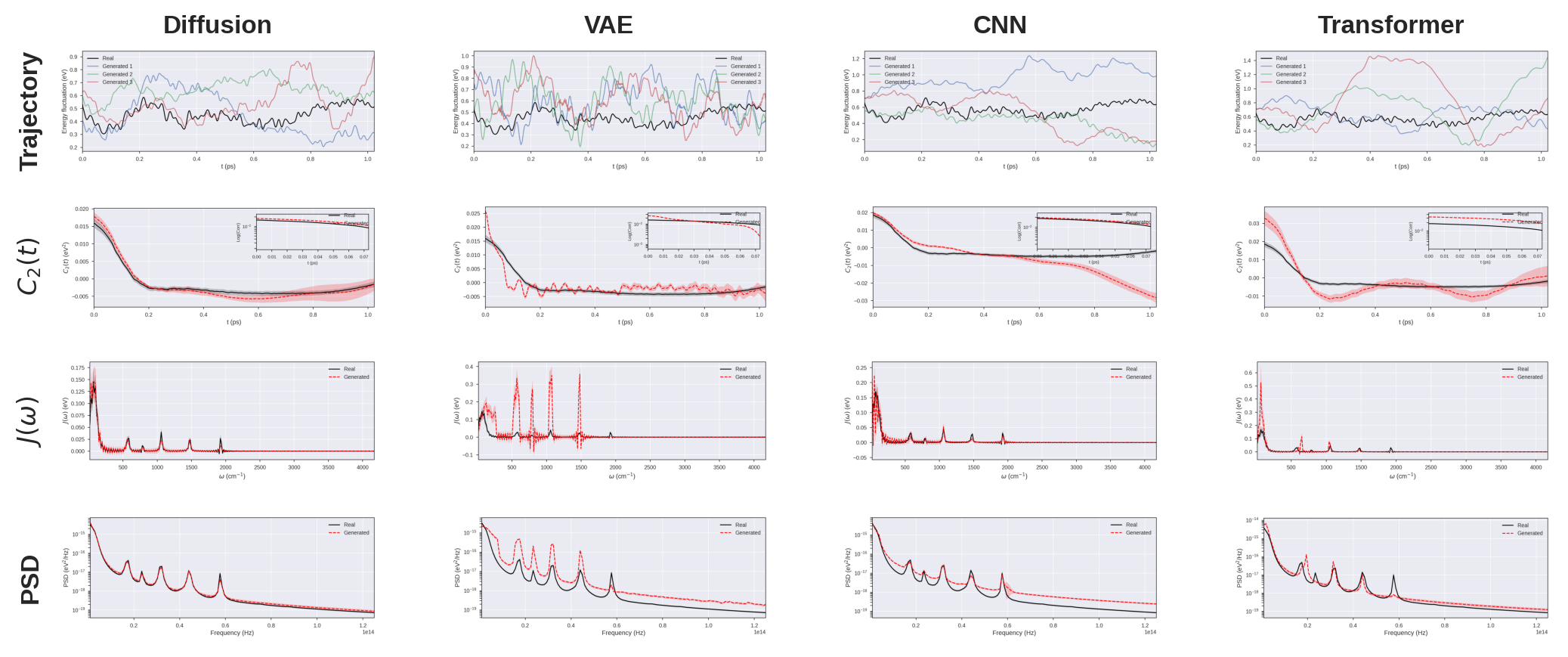}
  \caption{Generated statistics for ETG (one generation draw per model).
  Layout, colours, and ensemble sizes as in Fig.~\ref{fig:data_expnoise}.}
  \label{fig:data_peptoid}
\end{figure*}

ETG is the harder target, with a non-Gaussian amplitude distribution and
spectral structure on several timescales. The ranking from the OU process
nevertheless carries over (Table~\ref{tab:crossmodel_peptoid},
Fig.~\ref{fig:data_peptoid}). The diffusion model is best on every metric except
KS$_{\mathrm{amp}}$, on which the VAE is better, as for the OU process, and
reproduces the positions and relative heights of the spectral peaks.
The VAE matches the amplitude distribution but misses the increment distribution: it learns how the values are distributed without learning how they are correlated in time.
The autoregressive models show the opposite failure, reproducing the shape of $C_2(\tau)$ but with an amplitude distribution that is too narrow --- the same regression to the conditional mean seen in the white-noise limit of the OU process.

\begin{table*}[t]
  \centering
  \caption{Error metrics for ETG
  (Eqs.~\eqref{eq:mse_metric}--\eqref{eq:ks_metric}; best configuration per
  model). Each entry is the median of five generation draws from the same
  trained model, with only the sampling seed varied (seeds as in
  Table~\ref{tab:diffsweep}), so that the diffusion row coincides with the
  default of Table~\ref{tab:diffsweep}. Best per column in bold.}
  \label{tab:crossmodel_peptoid}
  \begin{tabular}{lrrrrr}
\toprule
\textbf{Method} & \textbf{KS$_\mathrm{amp}$} & \textbf{KS$_\mathrm{inc}$} & \textbf{MSE $C_2$} & \textbf{MSE $J$} & \textbf{MSE logPSD} \\
\midrule
\textbf{Diffusion} & $1.30\times10^{-2}$ & {\boldmath$1.37\times10^{-2}$} & {\boldmath$4.15\times10^{-7}$} & {\boldmath$2.47\times10^{-6}$} & {\boldmath$9.88\times10^{-4}$} \\
\textbf{VAE} & {\boldmath$5.70\times10^{-3}$} & $2.39\times10^{-1}$ & $6.09\times10^{-6}$ & $9.55\times10^{-4}$ & $3.06\times10^{-1}$ \\
\textbf{CNN} & $3.72\times10^{-1}$ & $6.47\times10^{-2}$ & $7.60\times10^{-5}$ & $8.09\times10^{-5}$ & $1.80\times10^{-1}$ \\
\textbf{Transformer} & $1.79\times10^{-1}$ & $4.55\times10^{-2}$ & $2.83\times10^{-5}$ & $2.11\times10^{-4}$ & $2.90\times10^{-2}$ \\
\bottomrule
\end{tabular}

\end{table*}

\medskip

Both tests show that a generative model trained on a small portion of the reference data can reproduce the correlation function, spectrum, and distributions of that reference.
They also reveal a structural difference between the model
classes. The diffusion model, which generates the whole trajectory at once,
is the most successful. The VAE also generates simultaneously but falls
short: it reproduces the distribution of values without their local correlations. The Transformer, by contrast, performs far better at large
$\tau$ than at small $\tau$, tracking how predictable the signal is. These
systematic differences are examined in Section~\ref{sec:discussion}.

\subsection{Diffusion Model hyperparameter study}
\label{sec:diffsweep}

\begin{figure}[tb]
  \centering
  \includegraphics[width=0.62\linewidth]{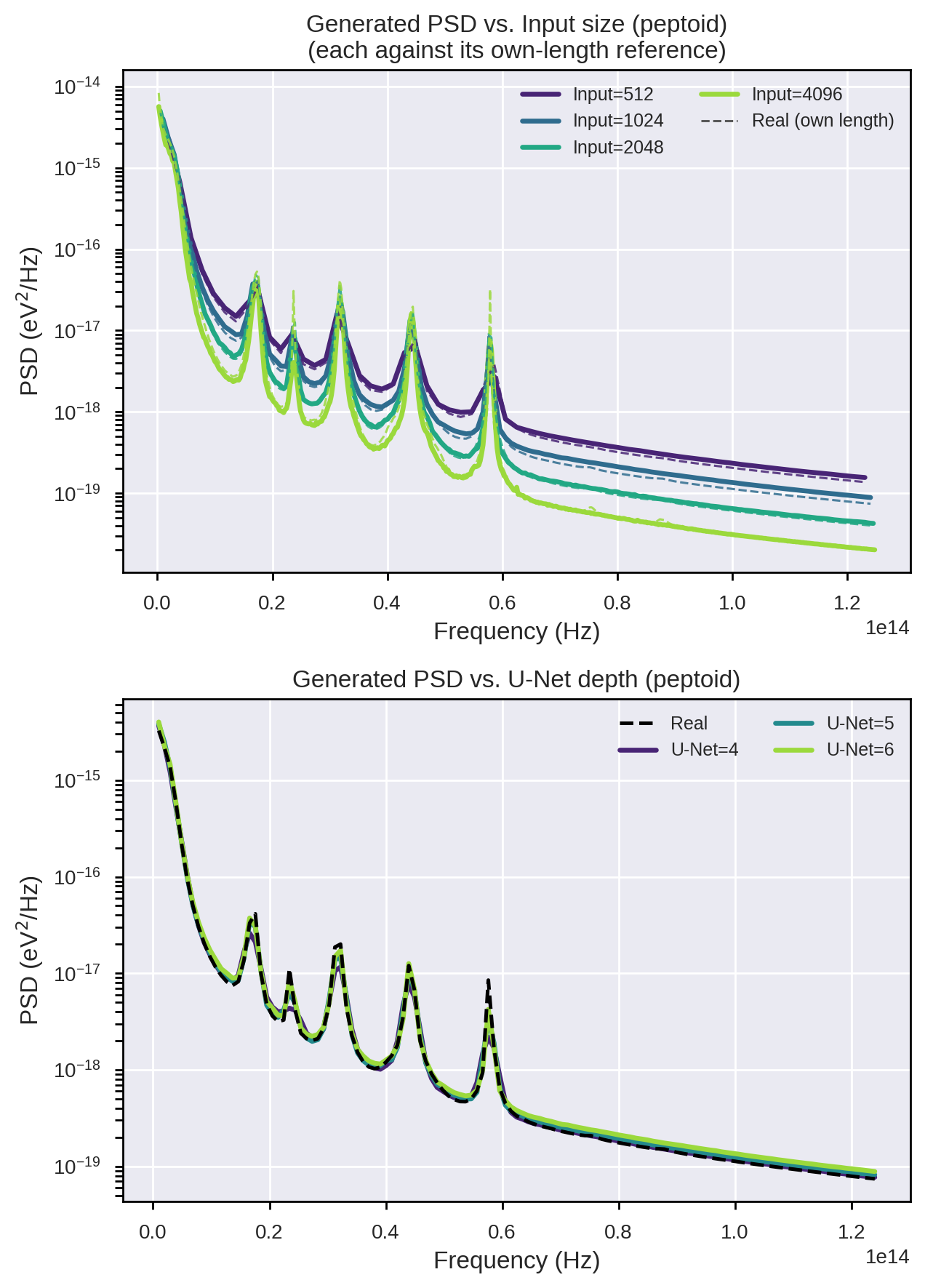}
  \caption{Generated PSD for ETG as a function of input length (top) and
  U-Net depth (bottom). Black: real reference. Each input-length panel is
  compared with a reference computed at the same length, because the
  periodogram itself depends on window length: leakage from the strong
  low-frequency peak dominates the high-frequency tail and decreases roughly
  as $1/L$, so offsets between input-length panels reflect the estimator, not
  model error. Spectra at different lengths also lie on different frequency
  grids ($256$ bins at $L=512$ to $2048$ at $L=4096$). Over the band
  $f<\nu_{\mathrm{Nyq}}/4$, where $\nu_{\mathrm{Nyq}}=1/(2\Delta t)$, length
  $2048$ gives the best match (MSE\,$\log_{10}$PSD $2.5\times10^{-3}$); in the
  high-frequency tail, length $4096$ does ($6.1\times10^{-4}$). Results at
  fixed length, including all cross-model comparisons, are unaffected. One
  representative draw is shown per configuration; Table~\ref{tab:diffsweep}
  reports medians over five.}
  \label{fig:diff_logpsd}
\end{figure}

Before turning to the third question, we check that the performance of the
diffusion model does not depend on a particular configuration. Three
hyperparameters are varied one at a time about the default of
Section~\ref{sec:protocol} (input length $1024$, stride $10$, 6 U-Net
layers): the input length, which is also the length of the generated
trajectories; the stride between training windows; and the U-Net depth
(Table~\ref{tab:diffsweep}, Fig.~\ref{fig:diff_logpsd}).

The dominant source of scatter is sampling rather than training. For a fixed
trained model, repeating only the reverse-diffusion step with new random
seeds changes MSE\,$\log_{10}$PSD by up to an order of magnitude, more than
most of the differences between configurations. Each entry of
Table~\ref{tab:diffsweep} is therefore a median over five generation draws,
and a difference is counted as real only if it exceeds the spread of the
default's own draws. By this criterion the default is best or tied-best on
every metric.

The stride sets the number of training samples through
Eq.~\eqref{eq:n_windows}.
Overlapping windows carry no additional information about the process, but they are not redundant for the model. The networks are not invariant to time shifts, so a window displaced by a single point is a distinct training example.
Overlap also supplies the number of examples that a network of
$\sim\!10^{7}$ parameters (Table~\ref{tab:cost}) requires, and exposes
it to the stationarity of the process. Strides of $10$ and $20$ are indistinguishable, while denser overlap
($s=2$, $5$) is measurably worse.

Depth shows the same plateau: five and six layers are indistinguishable,
while a four-layer network is measurably worse. Input length matters more.
The default outperforms all other lengths over the full frequency band and on
the amplitude distribution, although longer inputs reproduce parts of the
spectrum better (Fig.~\ref{fig:diff_logpsd}). The cross-model
comparison was thus carried out near the optimum, and its conclusions do not
depend on tuning.

\begin{table*}[t]
  \centering
  \caption{Diffusion hyperparameter sweep for ETG, varying one parameter at a
  time about the default (``def''). Each entry is the median of five
  generation draws from the same trained model, with only the sampling seed
  varied. The default's own draws span $3.8\times10^{-4}$ to
  $1.7\times10^{-3}$ in MSE\,$\log_{10}$PSD; a median within this range is
  counted as a tie. Best per column within each group in bold.}
  \label{tab:diffsweep}
  \begin{tabular}{lrrrr}
\toprule
\textbf{value} & \textbf{KS$_\mathrm{amp}$} & \textbf{MSE $C_2$} & \textbf{MSE $J$} & \textbf{MSE logPSD} \\
\midrule
\multicolumn{5}{l}{\textbf{Input size}} \\
512 & $2.46\times10^{-2}$ & $4.97\times10^{-7}$ & {\boldmath$1.20\times10^{-6}$} & $5.64\times10^{-3}$ \\
1024\,(def) & {\boldmath$1.30\times10^{-2}$} & {\boldmath$4.15\times10^{-7}$} & $2.47\times10^{-6}$ & {\boldmath$9.88\times10^{-4}$} \\
2048 & $1.40\times10^{-1}$ & $5.55\times10^{-7}$ & $3.34\times10^{-6}$ & $1.47\times10^{-2}$ \\
4096 & $6.72\times10^{-2}$ & $1.31\times10^{-6}$ & $8.58\times10^{-6}$ & $3.92\times10^{-3}$ \\
\midrule
\multicolumn{5}{l}{\textbf{Window stride}} \\
2 & $1.23\times10^{-1}$ & {\boldmath$3.90\times10^{-7}$} & $2.75\times10^{-6}$ & $3.49\times10^{-3}$ \\
5 & $1.16\times10^{-1}$ & $4.71\times10^{-7}$ & $3.10\times10^{-6}$ & $1.52\times10^{-2}$ \\
10\,(def) & {\boldmath$1.30\times10^{-2}$} & $4.15\times10^{-7}$ & {\boldmath$2.47\times10^{-6}$} & {\boldmath$9.88\times10^{-4}$} \\
20 & $1.57\times10^{-2}$ & $7.48\times10^{-7}$ & $3.31\times10^{-6}$ & $1.25\times10^{-3}$ \\
\midrule
\multicolumn{5}{l}{\textbf{U-Net depth}} \\
4 & $1.84\times10^{-2}$ & $7.06\times10^{-7}$ & $5.10\times10^{-6}$ & $3.84\times10^{-3}$ \\
5 & $2.55\times10^{-2}$ & $9.15\times10^{-7}$ & $4.44\times10^{-6}$ & $1.06\times10^{-3}$ \\
6\,(def) & {\boldmath$1.30\times10^{-2}$} & {\boldmath$4.15\times10^{-7}$} & {\boldmath$2.47\times10^{-6}$} & {\boldmath$9.88\times10^{-4}$} \\
\midrule
\bottomrule
\end{tabular}

\end{table*}

\subsection{Robustness across peptoid conformations}
\label{sec:structures}

\begin{figure*}[t]
  \centering
  \includegraphics[width=\linewidth]{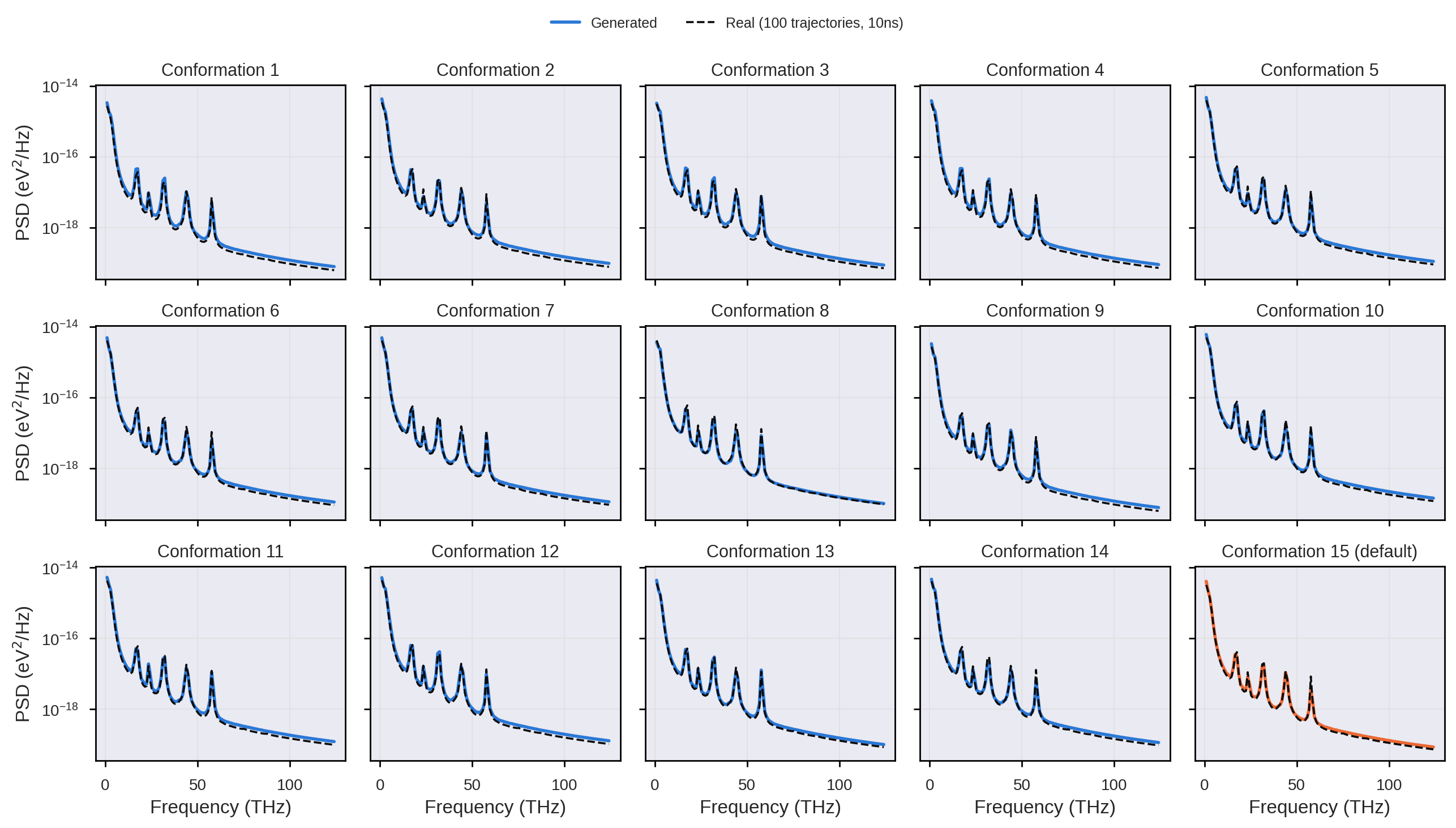}
  \caption{Generated and real PSD for the default diffusion model, trained
  and evaluated independently on each of the $15$ peptoid conformations
  (one representative draw; real curve from the $100$-trajectory reference).
  Orange: conformation $15$, used throughout the rest of the paper.}
  \label{fig:structures_psd}
\end{figure*}

All ETG results above use a single conformation of the peptoid scaffold
(conformation $15$). Our earlier rate study required $C_{UU}(t)$ for $15$
conformations (Section~\ref{sec:sys_etg}), so we ask whether the default
diffusion model, configured on conformation $15$ alone, performs equally
well on the other $14$ without retuning. The model is trained independently
on each conformation under the protocol of Section~\ref{sec:protocol}, and
each is scored against its own $100$-trajectory reference
(Table~\ref{tab:structures}, Fig.~\ref{fig:structures_psd}).

Accuracy is uniform across conformations. The spread of each metric across
the $15$ conformations is comparable to or smaller than the sampling scatter
of a single conformation (Section~\ref{sec:diffsweep}), and the peak
structure of the PSD is reproduced equally well throughout. Conformation $15$
ranks in the lower half on three of the four metrics, so the results
reported elsewhere are not a favorable special case.

Since each conformation was simulated independently, these results also show that the architecture and hyperparameters chosen on one conformation transfer to the others without retuning.

\begin{table*}[t]
  \centering
  \caption{Default diffusion model across the $15$ peptoid conformations.
  Each entry is the median of five generation draws, scored against the
  conformation's own reference ($100$ trajectories, $10$~ns). $\sigma$:
  trajectory standard deviation. NMSE: error in $C_2$ and $J(\omega)$
  normalized by the reference's mean square, since both scale as $\sigma^{2}$.
  $^{\ast}$Conformation $15$, used elsewhere in the paper, ranks $11$th,
  $10$th, $9$th, and $6$th of $15$ on KS$_{\mathrm{amp}}$, NMSE\,$C_2$,
  NMSE\,$J$, and MSE\,$\log_{10}$PSD.}
  \label{tab:structures}
  \begin{tabular}{lrrrrr}
\toprule
\textbf{Conf.} & \textbf{$\sigma$ (kJ/mol)} & \textbf{KS$_\mathrm{amp}$} & \textbf{NMSE $C_2$} & \textbf{NMSE $J$} & \textbf{MSE logPSD} \\
\midrule
1 & 16.8 & $1.57\times10^{-2}$ & $2.33\times10^{-2}$ & $5.57\times10^{-2}$ & $2.24\times10^{-3}$ \\
2 & 16.3 & $2.98\times10^{-2}$ & $3.12\times10^{-2}$ & $4.84\times10^{-2}$ & $1.67\times10^{-3}$ \\
3 & 16.1 & $2.01\times10^{-2}$ & $2.24\times10^{-2}$ & $6.92\times10^{-2}$ & $1.16\times10^{-3}$ \\
4 & 16.5 & $2.62\times10^{-2}$ & $3.19\times10^{-2}$ & $9.88\times10^{-2}$ & $1.83\times10^{-3}$ \\
5 & 16.2 & $2.73\times10^{-2}$ & $2.39\times10^{-2}$ & $2.68\times10^{-2}$ & $1.18\times10^{-3}$ \\
6 & 15.9 & $2.78\times10^{-2}$ & $2.61\times10^{-2}$ & $2.76\times10^{-2}$ & $9.66\times10^{-4}$ \\
7 & 16.4 & $1.31\times10^{-2}$ & $1.89\times10^{-2}$ & $2.26\times10^{-2}$ & $1.03\times10^{-3}$ \\
8 & 16.6 & $2.88\times10^{-2}$ & $3.31\times10^{-2}$ & $4.80\times10^{-2}$ & $1.58\times10^{-3}$ \\
9 & 15.7 & $1.21\times10^{-2}$ & $3.40\times10^{-2}$ & $3.04\times10^{-2}$ & $1.32\times10^{-3}$ \\
10 & 18.6 & $3.71\times10^{-2}$ & $2.04\times10^{-2}$ & $1.92\times10^{-2}$ & $1.11\times10^{-3}$ \\
11 & 18.0 & $1.42\times10^{-2}$ & $2.37\times10^{-2}$ & $3.09\times10^{-2}$ & $1.46\times10^{-3}$ \\
12 & 17.8 & $4.38\times10^{-2}$ & $2.03\times10^{-2}$ & $2.94\times10^{-2}$ & $1.83\times10^{-3}$ \\
13 & 18.0 & $2.36\times10^{-2}$ & $2.45\times10^{-2}$ & $2.06\times10^{-2}$ & $7.85\times10^{-4}$ \\
14 & 16.8 & $3.82\times10^{-2}$ & $1.75\times10^{-2}$ & $3.32\times10^{-2}$ & $1.14\times10^{-3}$ \\
15$^\ast$ & 15.8 & $2.97\times10^{-2}$ & $2.55\times10^{-2}$ & $3.15\times10^{-2}$ & $1.14\times10^{-3}$ \\
\midrule
\textbf{median} & 16.5 & $2.73\times10^{-2}$ & $2.39\times10^{-2}$ & $3.09\times10^{-2}$ & $1.18\times10^{-3}$ \\
\bottomrule
\end{tabular}

\end{table*}

\subsection{Convergence with sample size}
\label{sec:convergence}

\begin{figure*}[t]
  \centering
  \includegraphics[width=0.32\linewidth]{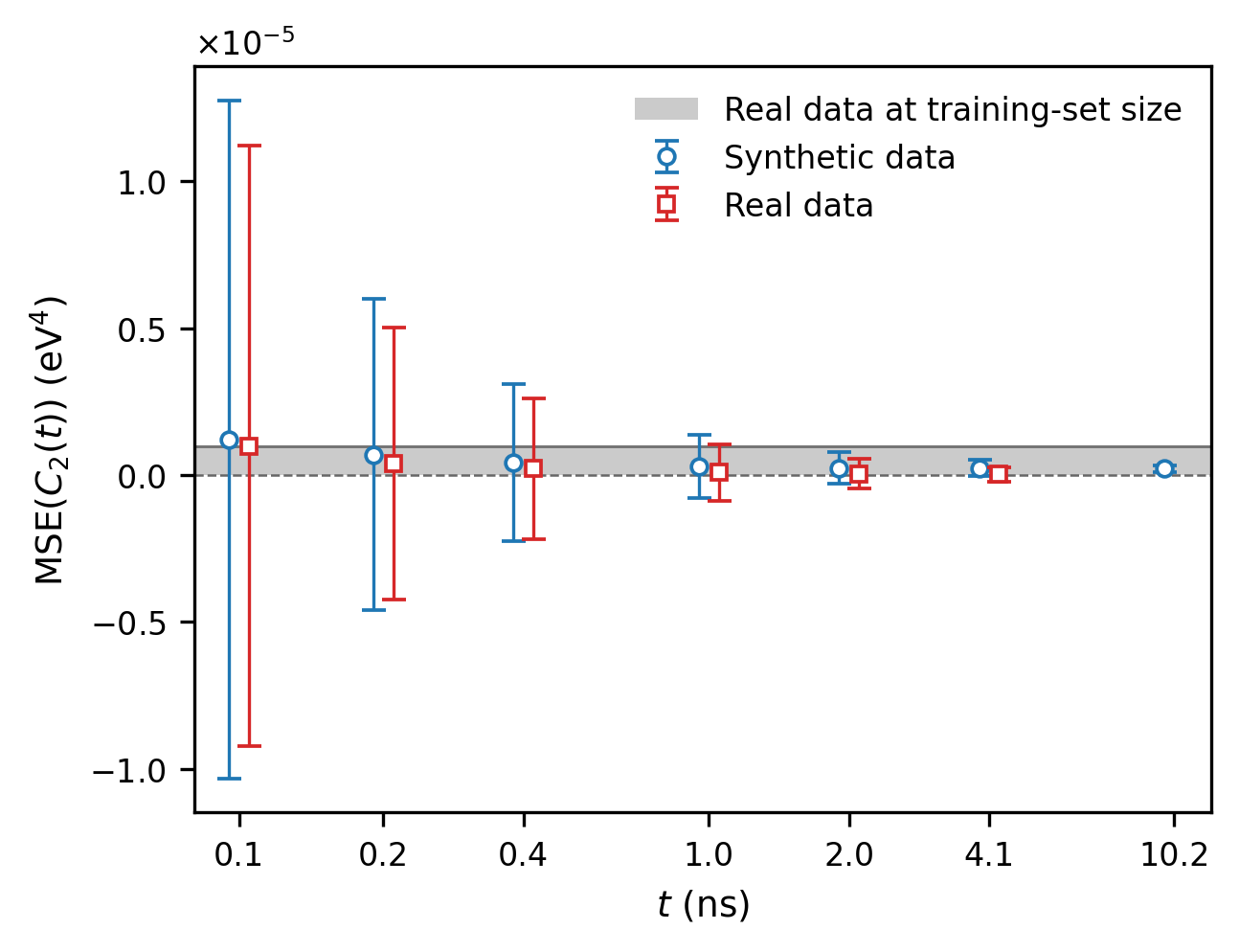}\hfill
  \includegraphics[width=0.32\linewidth]{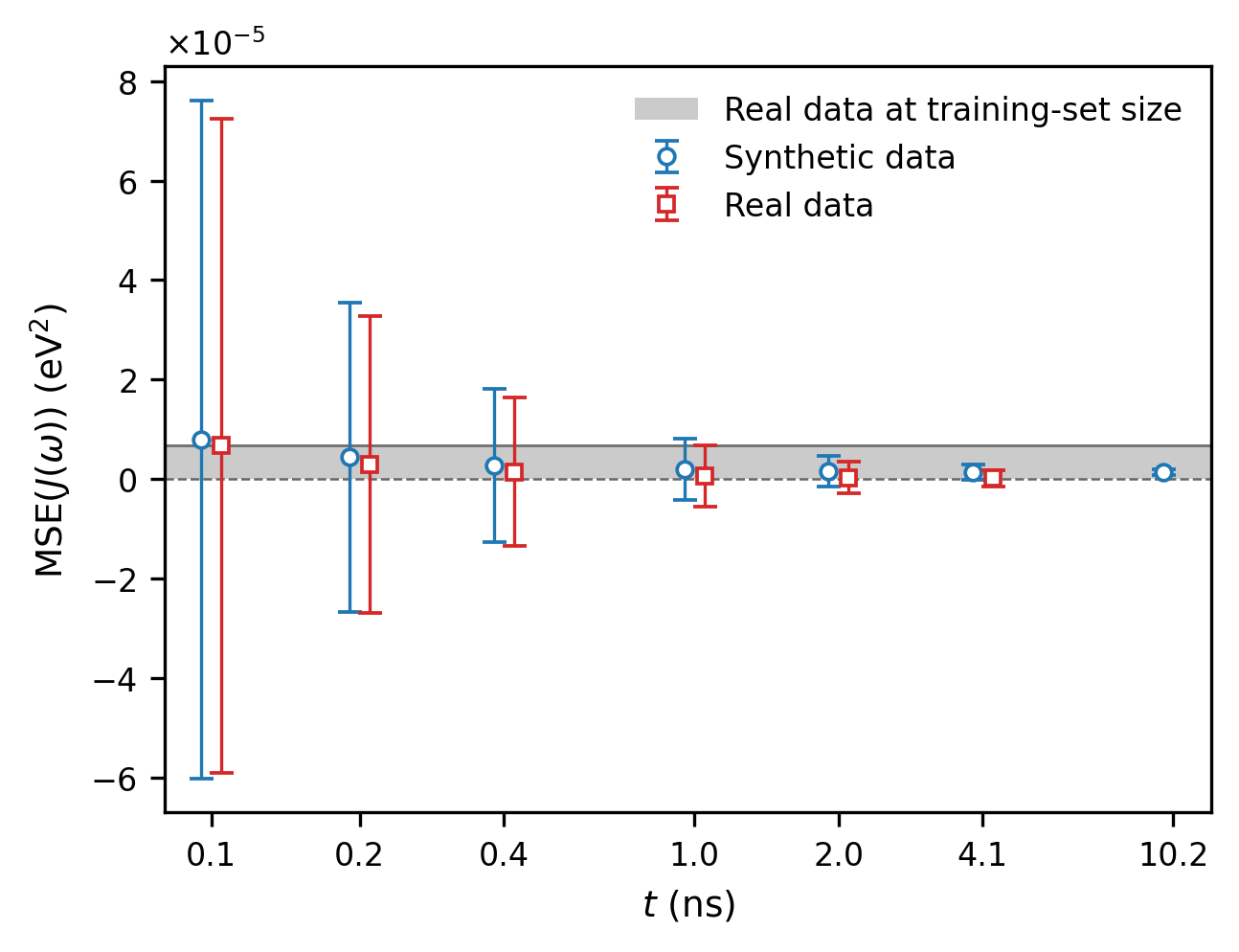}\hfill
  \includegraphics[width=0.32\linewidth]{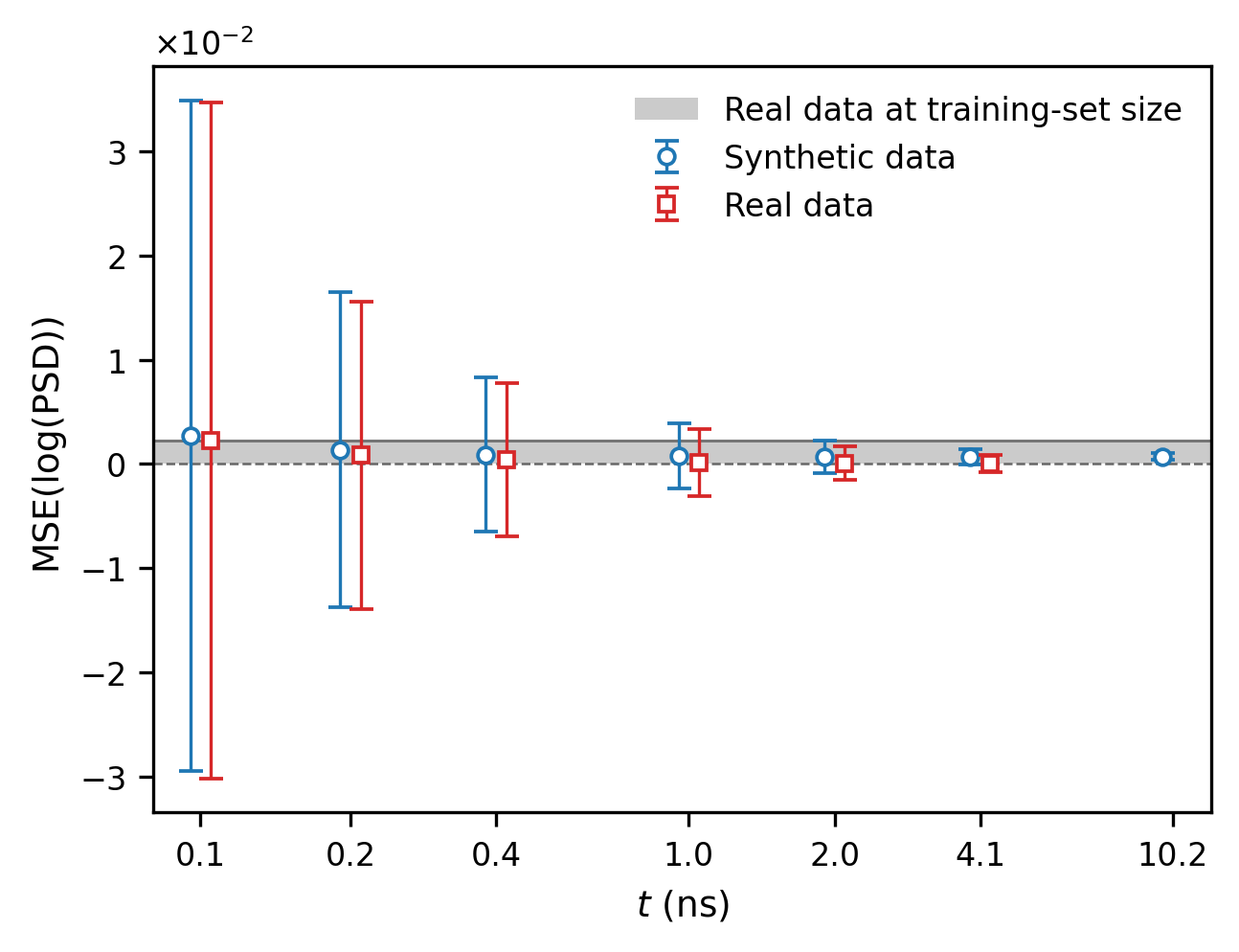}\\[2pt]
  \includegraphics[width=0.32\linewidth]{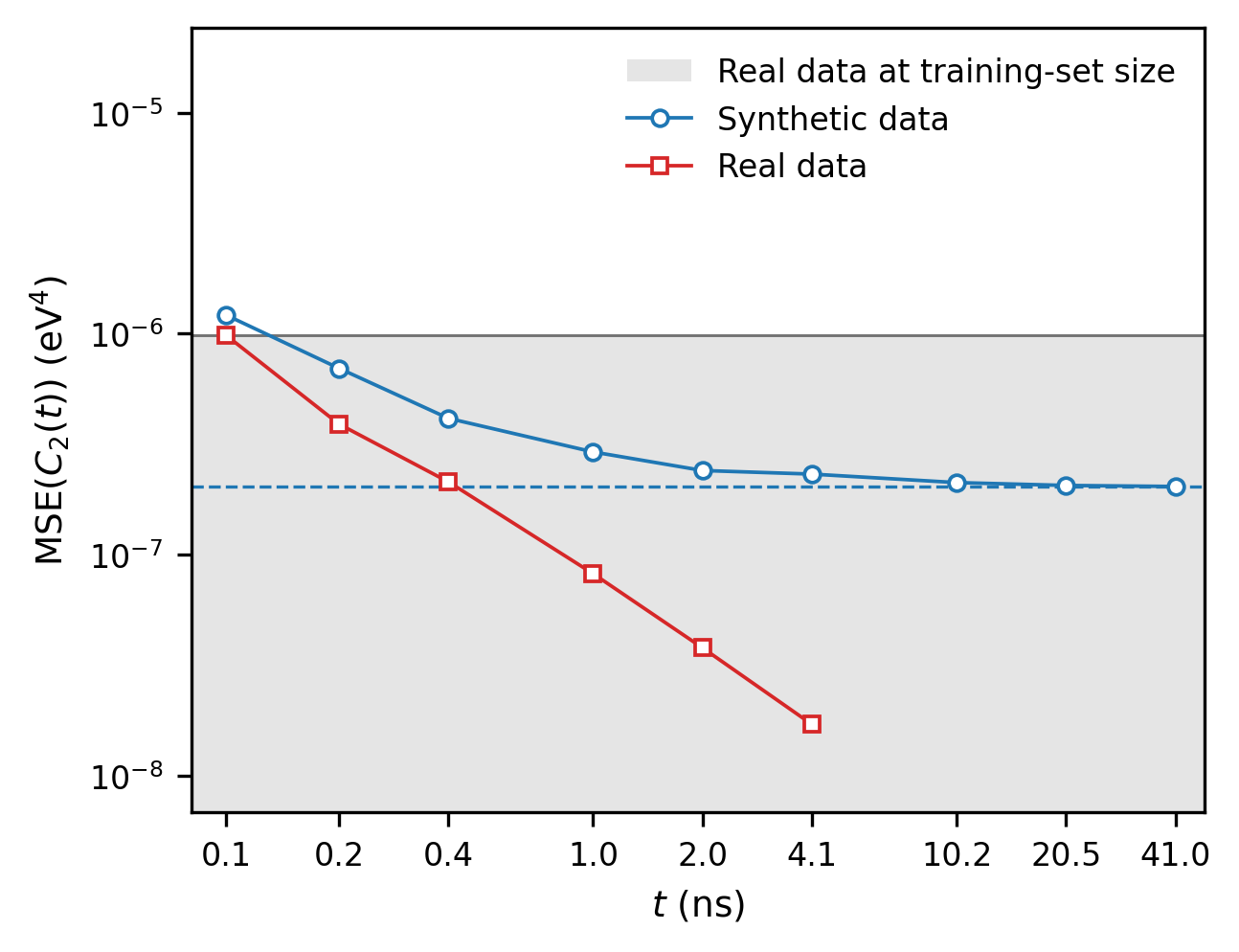}\hfill
  \includegraphics[width=0.32\linewidth]{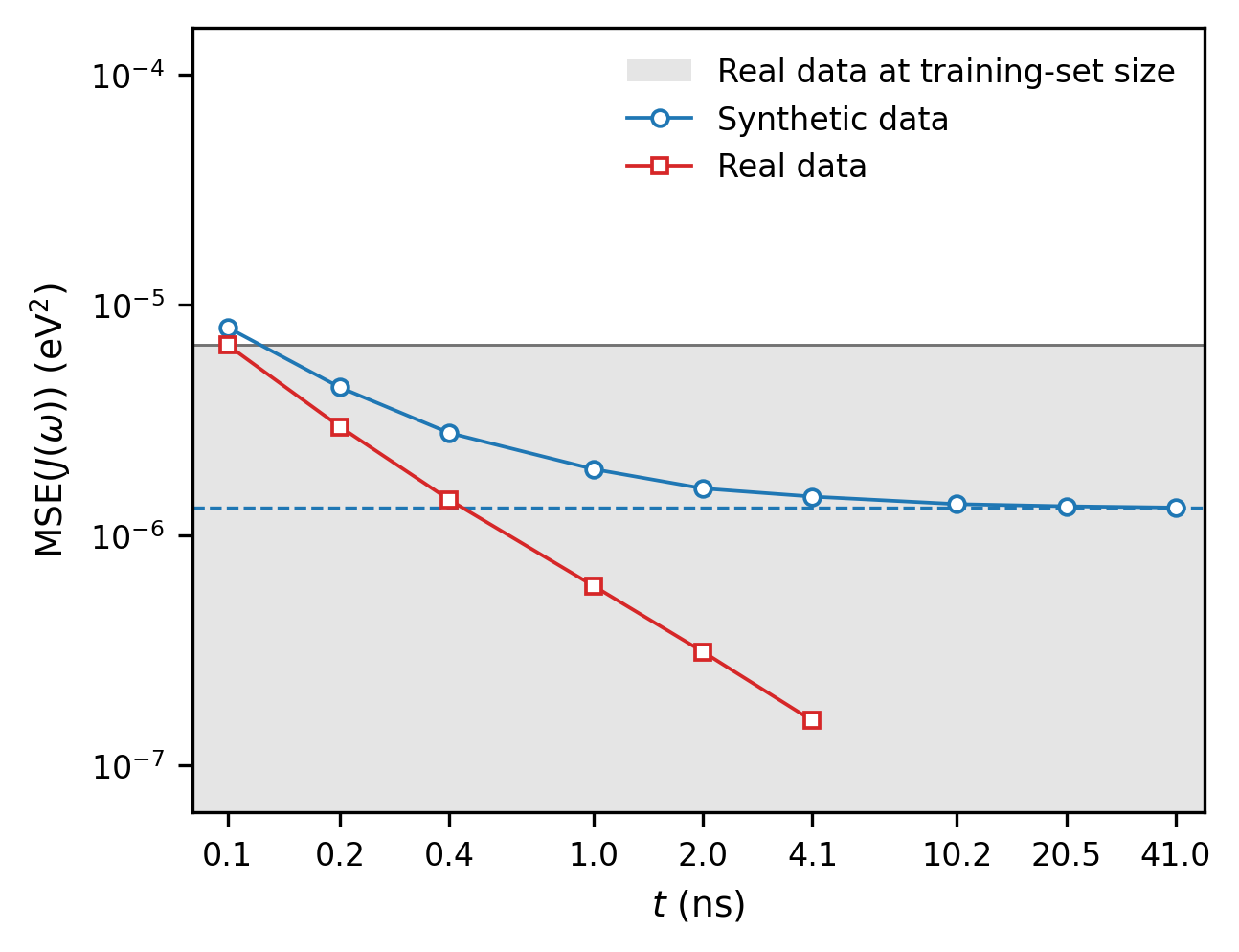}\hfill
  \includegraphics[width=0.32\linewidth]{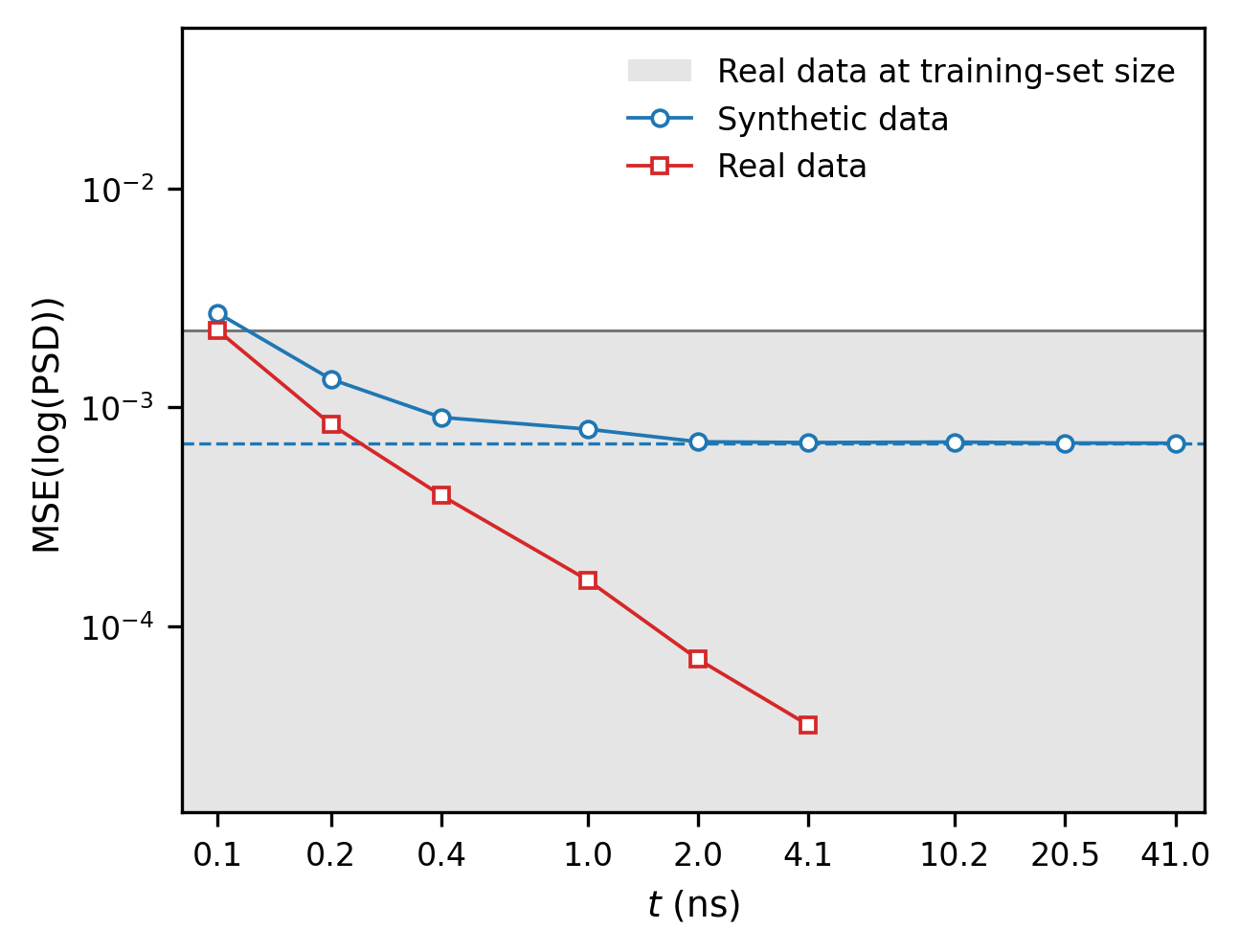}
  \caption{Error of the estimated statistics with respect to the converged
  reference ($100$ trajectories, $10$~ns) versus equivalent simulation time
  $t=N\times1.024$~ps, for ETG conformation $15$. Columns: $C_2(\tau)$,
  $J(\omega)$, and $\log_{10}$PSD. Top row: linear axes with error bars, $t\le10.2$~ns; bottom row: log--log axes without
  error bars, extended to $41$~ns. Points and error bars are defined in
  Supporting Information, Section~4. Blue: synthetic data from the default
  diffusion configuration. Red: real data. Grey band: real-data MSE at the
  training-set size ($0.1$~ns). Dashed blue line: converged synthetic MSE.}
  \label{fig:convergence}
\end{figure*}

This subsection addresses the third question: how much synthetic data can
improve an estimate, and what limits the improvement.
Figure~\ref{fig:convergence} shows the error of each statistic as a function
of the amount of data used to estimate it, expressed as equivalent simulation
time, for synthetic trajectories from the diffusion model and for real MD
trajectories. Both are scored against the converged reference, and the real
trajectories are disjoint from those that define it, so the two curves are
directly comparable. The grey band marks the error of the estimate from the
training data alone.

Two features answer the question. First, the synthetic estimate falls well
below the grey band: once converged, it is several times closer to the
reference than the training data it was learned from. Generative
augmentation therefore improves the statistics rather than merely
reproducing them. Second, the improvement is bounded. Beyond about $4$~ns of
synthetic data the error stops decreasing, whereas the error of real data
keeps falling and reaches the converged synthetic level at a crossover time
$t^{*}$. The crossover measures the value of the generator in units of
simulation time. Across four models trained on the same $0.1$~ns
trajectory and differing only in their random initialization (training
seed), a generator
trained on $0.1$~ns of MD gives estimates as accurate as $0.25$--$0.50$~ns of
MD, $2.5$ to $5$ times its training data. The spread comes mainly from the
PSD, whose converged level varies between models through a small offset in
the high-frequency tail.

\subsection{Computational cost and model capacity}
\label{sec:cost}

A difference in performance between models of different size could reflect
capacity rather than model class, since performance generally improves with
model size when data are plentiful~\cite{Kaplan2020,Hoffmann2022}. Here the
training data are fixed and limited, so model size must be matched to them:
too small a network cannot represent the statistics, and too large a network
overfits the single training trajectory. The size of each model was therefore
optimized for the present training set (Supporting Information; parameter
counts in Table~\ref{tab:cost}), and each is compared at its optimum. The
differences between classes are not a matter of capacity. Within the
diffusion model, a sixteenfold reduction in parameter count changes the error
by only a factor of a few (Section~\ref{sec:diffsweep}), far less than the
differences between classes.

Table~\ref{tab:cost} lists the wall-clock cost of training each model and
generating $1000$ trajectories, together with the five- and four-layer
diffusion variants of Section~\ref{sec:diffsweep}. Once trained, the
simultaneous models generate quickly, the VAE in a single forward pass,
whereas the Transformer is by far the most expensive, because its rollout is
strictly sequential. Among the diffusion variants, five layers matches the
accuracy of the default at lower cost, while four layers is cheaper still but
less accurate. The MD cost is an order-of-magnitude estimate, included only
for scale: trajectories that take many CPU-hours to simulate take seconds to
generate. The per-trajectory comparison overstates the saving, however.
Because synthetic data stop improving the estimate beyond the plateau of
Section~\ref{sec:convergence}, the relevant comparison is the cost of training
and generating up to that plateau against the cost of the additional MD that
reaches the same accuracy.

The approach also parallelizes in a way MD does not.
For a fixed trained generator, synthetic trajectories are conditionally
independent draws and can be generated in a single batched pass;
unlike MD, simultaneous generation is also parallel along time
within each trajectory, whereas both MD and autoregressive rollout must
advance one step at a time.

\begin{table*}[t]
  \centering
  \caption{Computational cost for ETG (best configuration per model, plus the
  five- and four-layer diffusion variants of Section~\ref{sec:diffsweep};
  generation of $1000$ trajectories (1ns)). $^{\dagger}$MD times are rough,
  machine-dependent estimates on CPU, and not directly
  comparable with the GPU times of the models.}
  \label{tab:cost}
  \begin{tabular}{lrrr}\toprule
\textbf{Method} & \textbf{\# params} & \textbf{train (s)} & \textbf{generate (s)} \\
\midrule
\textbf{Diffusion (6 layers)} & 22,941,233 & 210 & 34.51 \\
\textbf{Diffusion (5 layers)} & 5,788,209 & 163 & 23.09 \\
\textbf{Diffusion (4 layers)} & 1,471,537 & 137 & 16.52 \\
\textbf{VAE} & 1,672,449 & 40 & 0.01 \\
\textbf{CNN} & 4,279,169 & 35 & 2.43 \\
\textbf{Transformer} & 463,105 & 440 & 250.10 \\
\midrule
\textbf{ETG(MD) (real)}$^\dagger$ & --- & --- & $\sim\!3\times10^{4}$ \\
\textbf{OU (real)}$^\dagger$ & --- & --- & $\sim\!10^{0}$ \\
\bottomrule
\end{tabular}

\end{table*}

\subsection{Summary}
\label{sec:results_summary}

In this section we conclude by revisiting the three questions posed in the Introduction. First, the
statistics of equilibrium fluctuations can be learned from a sample too small
to converge them: the diffusion model, trained on one hundredth of the
reference data, reproduces the amplitude distribution, correlation function,
and spectrum of both data sources (Section~\ref{sec:crossmodel}), with a single
configuration across $15$ independently simulated conformations
(Section~\ref{sec:structures}). Second, the choice of model matters, and along
a structural line. The diffusion model, which generates whole trajectories, is
accurate at every correlation time; the sequential models are competitive only
when the signal is strongly correlated and otherwise regress to the
conditional mean; and the VAE reproduces the distribution of values but not
their temporal correlations. These differences reflect neither tuning nor
capacity (Sections~\ref{sec:diffsweep} and~\ref{sec:cost}). Third, the
improvement is real but bounded. The synthetic estimate is closer to the
converged reference than the training data alone, but it levels off at a
quality floor, so augmentation is worth a fixed multiple of the training data
rather than an unlimited amount of it (Section~\ref{sec:convergence}).

\section{Discussion}
\label{sec:discussion}

The Results section answer the three questions posed in the Introduction. In this
section we examine why the answers take the form they do: the mechanism behind
the ordering of the models, the relation of the present problem to
machine-learned MD, the limits of augmentation, and the
directions in which the approach can be extended.

\subsection{The deterministic--stochastic asymmetry}
\label{sec:disc_asym}

The ordering of the models follows from what each is trained to estimate. A
sequential model trained on the squared error of the next value learns the
conditional mean $\mathbb{E}[x_{t+1}\mid x_{\le t}]$. When the next value is
nearly determined by the past, as for the OU process at long $\tau$, the
conditional mean is an adequate generator. When it is not, the rollout
returns the mean instead of a sample and the generated ensemble collapses, as
observed at short $\tau$ and in the under-dispersed ETG amplitudes. No
increase in capacity repairs this, because the failure lies in the estimand
rather than in its approximation. A simultaneous model instead learns the
joint distribution of the whole trajectory, and its error enters once,
integrated over the diffusion process, rather than compounding along the
trajectory (Appendix~\ref{sec:arch_formal}). The collapse to the mean is
specific to the point-prediction outputs used here: a sequential model with a
probabilistic output would sample rather than average, but it would remain
subject to the compounding of per-step error along the trajectory.

The same boundary appears in machine-learned MD.
Deterministic propagators, which regress the next configuration, reach
strides of only one to two orders of magnitude beyond the MD timestep and
typically require a thermostat during generation, which supplies the
stochasticity their deterministic map cannot. Transition-density models,
which learn a conditional distribution, reach strides several orders of
magnitude larger. That a separate community,
working with atomistic configurations rather than scalar observables, meets
the same boundary indicates that it is structural rather than specific to our
test systems.

\subsection{Relation to machine-learned molecular dynamics}
\label{sec:mlmd_differences}

The present problem is nevertheless not a special case of trajectory
surrogacy. Learned propagators are judged on pathwise fidelity --- whether a
generated trajectory is a trajectory of the system --- hence their emphasis on
conservation laws, time-reversal structure, and Chapman--Kolmogorov
consistency. For a stochastic observable such as $U(t)$, no single
realization carries physical meaning. The physics resides in the ensemble,
and two ensembles with identical statistics are equivalent for every purpose
the observable serves. Pathwise fidelity is therefore sufficient but not
necessary, and relaxing it removes the constraints that make trajectory
surrogacy difficult: there is no conservation law to impose and no rollout to
stabilize, and whole trajectories can be generated at once. The price is a
narrower scope. The approach produces no configurations and cannot answer the
structural or mechanistic questions that trajectory surrogates address.

The two approaches also differ in their data. Learned propagators are trained
on large corpora assembled in advance and amortize that cost across the many
systems to which they transfer. The generator used here is fitted only to the
trajectories of the system under study, with no pre-training, and must repay
its cost within that system. Which approach is appropriate is decided by the
estimand: a configuration or mechanism calls for a trajectory surrogate, a
statistical property of a stochastic observable for a generator of the kind
studied here.

\subsection{Limits of augmentation}
\label{sec:disc_limits}

Because the generator is trained on the same trajectories used for
estimation, the data processing inequality applies in its tightest form,
Eq.~\eqref{eq:dpi_bound} with $N_t=N_r$: synthetic trajectories carry no
information about the system beyond that of the real sample. The benefit of
augmentation therefore cannot be an information gain; it arises because the
trained generator acts as a regularized estimator of the real data, trading
variance for bias. The error of an estimate from a trained generator
separates into a sampling part, which additional synthetic samples remove,
and a part fixed by the generator itself --- its bias and the uncertainty it
inherits from a finite training set --- which they cannot
(Eq.~\eqref{eq:error_decomp}). The plateau of Section~\ref{sec:convergence}
is this floor observed directly. A quantitative theory of the floor --- its
dependence on the size of the training set and on the fidelity of the
generator, and hence the number of synthetic samples worth drawing ---
requires generators trained on independent training sets, and is left to
future work.

\subsection{Further directions: beyond augmentation}
\label{sec:further}

The preceding sections have treated the trained generator as a means of
augmenting data, and have shown that in this role its value is bounded by the
information in the training trajectories. Augmentation is, however, the
simplest use of such a model rather than its only one. What has been trained
is a model of the law of a stochastic observable, which can be sampled cheaply
and conditioned on partial information, and these capabilities support
applications that a larger sample alone cannot.

One potential application is a learned bath for quantum dynamics and spectroscopy.
Several methods for condensed-phase quantum dynamics and spectroscopy treat
the environment as a classical stochastic process, in which fluctuating
energies drawn from MD enter the quantum propagation directly; examples are
the numerical integration of the Schr\"odinger equation for excitonic and
two-dimensional spectra and stochastic Liouville approaches. These methods
average an exponential of a time-integrated fluctuation over many
trajectories, converge slowly, and draw their input from expensive
simulation.
A generator trained on MD can supply such trajectories cheaply and can in principle retain statistical structure that a Gaussian harmonic-bath model discards, provided the relevant observables are validated.
Two extensions are required: joint generation of several correlated
observables, such as the site energies of a multichromophore system, and a
correction for the detailed balance that a classical bath lacks.

The second powerful direction is to generate statistics of systems not yet simulated.
A generator conditioned on a descriptor of the system --- a conformation, a
donor--acceptor distance, a temperature, or a solvent --- and trained across
several systems could produce the statistics of conditions it has never seen.
This is the regime $N_t>N_r$ of Remark~\ref{rem:bootstrap}: the model carries
information external to the system under study, and the benefit is no longer
bounded by Eq.~\eqref{eq:dpi_bound}. For the peptoid system, it would replace
the simulation of a handful of representative conformations with a screen of
the conformational space. The risk is extrapolation to systems unlike those in
the training family, where the model can be confidently wrong. Conditioning on
absolute time instead of a system descriptor extends the same idea to
non-equilibrium dynamics, such as solvent relaxation after a change in donor
charge state, where the relevant correlation functions depend on two times.

A third direction is conditional generation from partial information.
A diffusion model can also generate conditioned on part of a trajectory.
Generating each new window conditioned on its overlap with the previous one
produces trajectories longer than the training window, relaxing the
resolution limit of Section~\ref{sec:disc_limits}, although correlations
longer than one window remain beyond what the model can learn. The same
mechanism fills gaps in incomplete or noisy trajectories, such as
experimental single-molecule traces. Guiding generation toward rare regions,
such as the gap crossings that dominate inverted-region rates, and
reweighting by the model likelihood would give access to tail estimands,
provided the model's tail is accurate --- the caveat raised in
Section~\ref{sec:disc_limits}.

\section{Conclusions}
\label{sec:conclusions}

We have asked whether a generative model trained on a portion of MD
trajectories too small to converge can supply
the missing statistical sampling, using an analytically tractable
Ornstein--Uhlenbeck process and donor--acceptor energy-gap trajectories from all-atom MD as test cases.
The results reduce to three messages.

First, the statistics of stochastic but correlated data can be learned from
limited data and used to extend the sampling, within a limit set by the data
themselves. A diffusion model trained on a single $0.1$~ns trajectory
reproduces the amplitude distribution, correlation function, and spectrum of
the converged reference, with one configuration across $15$ independently
simulated conformations, and its estimates are as accurate as those obtained
from $2.5$ to $5$ times as much MD. The gain is bounded. Synthetic samples
carry no Fisher information about the system beyond that of the real data ---
the Fisher information being the quantity that bounds, through the
Cram\'{e}r--Rao inequality, the precision of any estimate
(Appendix~\ref{sec:gda_theory}) --- so the improvement is a reduction of
variance paid for in bias. It levels off at a floor set by the bias of the
generator together with the uncertainty it inherits from the finite training
data, neither of which additional synthetic sampling can remove. How that
floor depends on the size of the training set and on the fidelity of the
generator is left to future work.

Second, the choice of generative model is decisive, and it follows a
structural line rather than a ranking by capacity. The diffusion model, which
generates whole trajectories at once, is accurate at every correlation time.
Sequential models trained to predict the next value are competitive only when
the signal is strongly correlated and otherwise regress to the conditional
mean, while the VAE reproduces the distribution of values but not their
temporal correlations.

Third, generative augmentation is an extension of sampling, not a replacement
for dynamics. The equations of motion are unchanged, and the approach
produces neither configurations nor mechanisms. What it offers is additional
statistical sampling of the correlation functions and spectral densities on
which rate theory depends, drawn from the trajectories one can afford to
propagate. This distinguishes it from machine-learned MD,
which must reproduce trajectories pathwise and relies on large pre-training
corpora, whereas the present generator is trained only on the data at hand.

Beyond augmentation, the same learned model can serve as a bath for quantum
dynamics and nonlinear spectroscopy, supplying realistic non-Gaussian
fluctuations where harmonic models are now used. Conditioned on the system or
on time, it can produce the statistics of conformations and state points not
yet simulated, and of non-equilibrium processes. And conditioned on partial data, it can extend trajectories
beyond the training window and fill gaps in incomplete ones. A generative
model of a stochastic observable is in this sense more than a device for
augmenting data: it is a learned, cheaply sampled representation of the
process itself.

\begin{acknowledgement}
This material is based upon work supported by the U.S. Department of Energy, Office of Science, Office of Basic Energy Sciences. BDD and EG were supported under Award No.~DE-SC0025404. MSC and BSH acknowledge support from FWP 83482 for implementing the molecular dynamics workflow and for generating and analyzing simulation data. MDB acknowledges support from FWP 77876 for parameterizing the porphyrins and analyzing the molecular dynamics data. This research used resources of the National Energy Research Scientific Computing Center (NERSC), a DOE Office of Science User Facility, using NERSC award BES-ERCAP0023738.
\end{acknowledgement}

\begin{suppinfo}
Model architectures and hyperparameters for the convolutional network, transformer,
variational autoencoder, and denoising diffusion probabilistic model; training
protocol, optimizer settings, and data preprocessing; and additional numerical
details supporting the results reported in Section~\ref{sec:results}.
\end{suppinfo}

\appendix

\vspace{0.5in}

\noindent
{\bf \LARGE APPENDIX}

\section{Information Bound and Error Decomposition}
\label{sec:gda_theory}

This appendix states the two results on which the main text relies: synthetic samples carry no information about the system beyond that contained in the data used to train the generator, and the error of an estimate computed from a trained generator separates into a part that additional synthetic samples remove and a part they cannot.
A quantitative theory of the latter --- its dependence on the
training-set size and on the fidelity of the generator, and hence the number of
synthetic samples worth drawing --- is left to future work.

\subsection{Setup and notation}
\label{sec:setup}

We write $P \equiv P_{\theta_0}$ for the true law of the data, indexed by the
parameter $\theta_0$ to be estimated, and $Q$ for the law realized by the
trained generator. The generator is trained on $N_t$ real samples, and
$N_s$ synthetic samples $\tilde x_1, \ldots, \tilde x_{N_s} \sim Q$ are drawn
from it. The score of the parametric family is
\begin{equation}
    s(x;\theta) = \frac{\partial}{\partial\theta}\log p(x;\theta),
    \label{eq:score_def}
\end{equation}
and $\mathcal{I}_P(\theta_0) = \mathbb{E}_P[s(x;\theta_0)^2]$ is the Fisher
information of a single real sample. We assume the standard regularity
conditions under which the Fisher information is well defined: the support of
$p(\cdot\,;\theta)$ does not depend on $\theta$, $p$ is twice continuously
differentiable in $\theta$, and differentiation and integration may be
exchanged.

\subsection{Synthetic samples information}
\label{sec:zero_info}

Once trained, the generator is a fixed object: a function of the trajectories
used to fit it, plus its own sampling noise. Drawing from it does not consult
the physical system again, much as enlarging a photograph of a spectrum yields
more pixels but no more resolution than the original exposure contained.
Formally, generation forms a Markov chain
$\theta \to X_{\mathrm{train}} \to Q \to \tilde{X}_{1:N_s}$, since the synthetic
samples are a randomized function of the training data with generator noise
independent of $\theta$. The data processing inequality for Fisher
information~\cite{Zamir1998} therefore bounds what the synthetic samples can
convey by what the training set contained,
\begin{equation}
    \mathcal{I}_{\tilde{X}_{1:N_s}}(\theta) \leq
    \mathcal{I}_{X_{\mathrm{train}}}(\theta) = N_t\,\mathcal{I}_P(\theta),
    \label{eq:dpi_bound}
\end{equation}
for any $N_s$. When the generator is trained on the same $N_r$ trajectories
used for estimation, as in this work, \eqref{eq:dpi_bound} holds with
$N_t = N_r$ and the Cram\'{e}r--Rao theorem~\cite{Cramer1946a, Rao1945} gives:

\begin{lemma}[Unbiased estimation is not improved by augmentation]
\label{lem:zero_info}
Any estimator $\hat\theta$ built from the real data and any number of
synthetic samples that is unbiased for $\theta_0$ satisfies
\begin{equation}
    \mathrm{Var}(\hat\theta) \geq \frac{1}{N_r\,\mathcal{I}_P(\theta_0)},
    \label{eq:crb_unbiased}
\end{equation}
independently of $N_s$.
\end{lemma}

Any benefit of augmentation therefore cannot be an information gain. It must
come from bias: the trained generator acts as a regularized estimator of the
real data, of the same kind as shrinkage estimators and smoothing fits, to
which Lemma~\ref{lem:zero_info} does not apply.

\subsection{Error decomposition for a generator}
\label{sec:error_decomp}

Let $\hat\theta$ be an estimate computed from $N_s$ samples of a generator
$Q$, which is itself random through its training data and its training
procedure. The law of total variance gives, without approximation,
\begin{equation}
    \mathbb{E}\big[(\hat\theta-\theta_0)^2\big]
    = \underbrace{\mathbb{E}_{Q}\big[\mathrm{Var}(\hat\theta \mid Q)\big]}_{\textnormal{sampling}}
    + \underbrace{\mathrm{Var}_{Q}\big[\mathbb{E}(\hat\theta \mid Q)\big]}_{\textnormal{training}}
    + \underbrace{\big(\mathbb{E}[\hat\theta]-\theta_0\big)^2}_{\textnormal{bias}} .
    \label{eq:error_decomp}
\end{equation}
Conditionally on $Q$ the synthetic samples are independent draws, so the
sampling term vanishes as $N_s \to \infty$ (as $1/N_s$ for a regular
estimator).

The remaining two terms reflect the variability of the fitted generator across training data and training procedure, together with its systematic bias. In the large-$N_s$ limit they define a floor that additional synthetic sampling cannot remove.
Because
$\mathbb{E}(\hat\theta \mid Q)$ is a function of the training data alone,
Lemma~\ref{lem:zero_info} bounds the training term below by
$1/(N_r\mathcal{I}_P)$ whenever $\mathbb{E}(\hat\theta \mid Q)$ is unbiased; a
floor below the error of the real data alone is possible only for a biased,
regularizing generator.

\begin{remark}[The empirical distribution as a generator]
\label{rem:empirical_q}
If $Q$ is the empirical distribution of the real data, synthetic sampling is
resampling, $\mathbb{E}(\hat\theta \mid Q)$ is the plug-in estimate from the
real data,
and as $N_s\to\infty$ the synthetic estimate converges to that plug-in estimate, so the floor is the error of the real data alone. Finite $N_s$ only adds resampling noise. Any improvement below this floor must come from the smoothing introduced by the fitted generator.
\end{remark}

\begin{remark}[Reading Fig.~\ref{fig:convergence}]
\label{rem:reading_fig}
For a single trained generator, increasing the number of synthetic samples
removes the sampling term, and the plateau of the synthetic curve is the
floor of \eqref{eq:error_decomp} for that generator. The four training seeds
of Section~\ref{sec:convergence} share one training trajectory, so their
spread samples the part of the training term due to the training procedure,
not the part due to the finite training data. Separating the two requires
generators trained on independent training sets.
\end{remark}

\begin{remark}[A replica must use external information]
\label{rem:bootstrap}
If a generator is fitted to the same $N_r$ samples used for estimation, then
by \eqref{eq:dpi_bound} it adds no information, as in the parametric
bootstrap~\cite{Efron1979}. A generator improves on the information in the
real data only when built from information beyond them: external training
data ($N_t > N_r$) or physical constraints on the form of the dynamics. This
is the regime occupied by transferable models pretrained on large external
corpora (Section~\ref{sec:discussion}).
\end{remark}

\section{Sequential and Simultaneous Generation on Stochastic Data}
\label{sec:arch_formal}
\label{sec:det_vs_stoch}

This appendix makes precise the two mechanisms invoked in the main text: the
failure of point-prediction autoregressive models on stochastic data, and the
absence of error compounding along the trajectory in score-based generation.
It also states what the comparison in this work does and does not test.

\subsection{Deterministic and stochastic conditional laws}

For a  ph{deterministic} process the next value is fixed by the history,
\begin{equation}
    p(x_t \mid x_{<t}) = \delta\big(x_t - f(x_{<t})\big),
    \label{eq:det_conditional}
\end{equation}
and generative modeling reduces to learning the map $f$. For a
 ph{stochastic} process the conditional law is a non-degenerate density: the
next value is distributed around its conditional mean with an innovation
variance $\mathbb{E}[\mathrm{Var}(x_t \mid x_{<t})] > 0$ that no amount of
history removes. The energy gap $U(t)$ is of this kind, since it is driven by
a bath of degrees of freedom that are not part of the observed history.

\subsection{Point-prediction autoregressive models}
\label{sec:ar_point}

The autoregressive models of this work output a single value and are trained
on its squared error, Eq.~\eqref{eq:loss_ar}. The minimizer of the squared
error is the conditional mean,
\begin{equation}
    f^{*}(x_{t-k+1}, \ldots, x_t) = \mathbb{E}[x_{t+1} \mid x_{t-k+1}, \ldots, x_t],
    \label{eq:cond_mean}
\end{equation}
so even a network of unlimited capacity learns the conditional mean and
nothing else. A rollout that feeds its own predictions back as input therefore
contains no innovation: the generated trajectory is a deterministic function
of its initial context. For deterministic data this is exact, since the
conditional mean is the map $f$ of \eqref{eq:det_conditional}. For stochastic
data it removes the fluctuations the model was meant to reproduce.

The OU process makes the consequence explicit. Its discretization,
$\xi_{n+1} = \rho\,\xi_n + h_n$ with $\rho = e^{-\Delta t/\tau}$ and $h_n$
independent of the past, has conditional mean $\rho\,\xi_n$, so the optimal
point-prediction rollout is
\begin{equation}
    \hat\xi_{n+m} = \rho^{m}\,\xi_n = e^{-m\Delta t/\tau}\,\xi_n .
    \label{eq:ou_rollout}
\end{equation}
The generated trajectory relaxes to zero on the correlation time, and the
variance of the generated ensemble decays as $\rho^{2m}$. When $\tau$ is
comparable to $\Delta t$ the collapse is complete within a few steps; when
$\tau$ is comparable to the $k$-point window it is only partial within the
window. This is the pattern of Table~\ref{tab:crossmodel_expnoise}, and it
arises from the training objective, not from limited capacity.

\subsection{Probabilistic autoregressive models}
\label{sec:ar_prob}

A sequential model that outputs the conditional distribution and samples from
it at each step does not suffer this collapse. For the OU process such a model
is exact: a first-order autoregressive model with Gaussian innovations
reproduces the discretization above. Trained by teacher forcing, with the
history drawn from the data, its objective is by the chain rule of the
Kullback--Leibler divergence
\begin{equation}
    \sum_{t=1}^{T} \mathbb{E}_{x_{<t}\sim P}\!\left[ D_{\mathrm{KL}}\big(p(\cdot\mid x_{<t}) \,\big\|\, q(\cdot\mid x_{<t})\big)\right],
    \label{eq:teacher_forcing}
\end{equation}
which is exact and additive over the horizon $T$. At generation time,
however, the history comes from the model's own output rather than from $P$.
A per-step error can then move the rollout into histories not seen in
training, and in the worst case the error grows as $O(\epsilon T^{2})$ for a
per-step error $\epsilon$~\cite{RossBagnell2010}. Whether this compounding
limits probabilistic autoregressive models on the systems studied here is not
tested in this work, which compares point-prediction models only.

\subsection{Score-based models}
\label{sec:score_based}

A score-based model does not factorize along the trajectory. It learns the
score $\nabla_x \log p_u(x)$ of noise-perturbed versions of the joint
distribution of a window, indexed by a diffusion time $u$, and generates by
integrating a reverse-time stochastic differential equation. With the
weighting $\lambda(u) = g(u)^{2}$, its error obeys~\cite{SongDurkanMurrayErmon2021}
\begin{multline}
    D_{\mathrm{KL}}(P \,\|\, Q) \leq \frac{1}{2}\int_0^{U_{\mathrm{d}}} \lambda(u)\,
    \mathbb{E}_{x \sim p_u}\!\left[\big\| s_\theta(x,u) - \nabla_x \log p_u(x)\big\|^2\right] du \\
    + D_{\mathrm{KL}}\big(p_{U_{\mathrm{d}}} \,\|\, \pi\big),
    \label{eq:song_bound}
\end{multline}
where $g(u)$ is the diffusion coefficient of the forward noising process,
$s_\theta$ is the learned score, $U_{\mathrm{d}}$ is the total diffusion time,
and the last term is the mismatch between the fully noised distribution and
the Gaussian prior $\pi$ from which generation starts, which decays
exponentially in $U_{\mathrm{d}}$. The learning error enters once, integrated
over diffusion time. There is no rollout along the trajectory and hence no
analogue of the horizon dependence of \eqref{eq:teacher_forcing}.

The bound requires a smooth density with a well-defined score. Stochastic data
provide one, because the innovation spreads the data over a full-dimensional
region. For deterministic data the distribution concentrates on a
lower-dimensional set, the score diverges near it, and training requires
substantial noise regularization~\cite{SongErmon2019}. This is why the
score-based model avoids the conditional-mean collapse of Section~\ref{sec:ar_point}. It does not establish a general ordering between simultaneous and probabilistic sequential generation;
as noted in Section~\ref{sec:ar_prob}, a probabilistic first-order autoregressive model represents the discretized OU process exactly. The fidelity of the trained
generator enters the bias term of the error decomposition in
Eq.~\eqref{eq:error_decomp}.

\begin{remark}[Scope of the architectural comparison]
\label{rem:ar_scope}
The failure of the sequential models reported in Section~\ref{sec:crossmodel}
is the collapse of Section~\ref{sec:ar_point}, a property of point prediction
under a squared-error objective. It establishes that such models are
unsuitable for stochastic data, not that sequential generation is. A
comparison with probabilistic sequential models, beginning with the exact
first-order autoregressive model of the OU process, is left to future work.
\end{remark}

\bibliography{refs,bib-eitan}

\end{document}


\maketitle

This Supporting Information collects the architectural specifications,
hyperparameters, and training protocols of the four generative models
compared in the main text, together with additional details on the
stability and reproducibility of diffusion sampling and the definition of
the error bars in the convergence study.

\section{Model Architectures}

Complete architectural specifications are given in Tables~\ref{tab:cnn},
\ref{tab:transformer}, \ref{tab:vae}, and~\ref{tab:ddpm}. All models were
implemented in PyTorch and trained on an NVIDIA RTX 5090 GPU. Every model
takes windows of $W=1024$ points as input (Section~5.1 of the main text);
the autoregressive models (CNN and Transformer) map such a window to the
next point, and the simultaneous models (VAE and DDPM) generate whole
$1024$-point windows. The parameter totals are those reported in Table~5 of
the main text.

\subsection{Convolutional neural network}

\begin{table}[H]
\centering
\caption{CNN architecture (autoregressive one-step predictor). Shapes are
(channels, length), with the batch dimension omitted. All convolutions use
kernel size $5$ and padding $2$; dropout after the flatten layer was set to
zero in the reported configuration.}
\label{tab:cnn}
\begin{tabular}{llll}
\toprule
\textbf{Layer} & \textbf{Input Shape} & \textbf{Output Shape} & \textbf{Parameters} \\
\midrule
Conv1D + ReLU & (1, 1024) & (32, 1024) & 192 \\
MaxPool1D (size 2) & (32, 1024) & (32, 512) & 0 \\
Conv1D + ReLU & (32, 512) & (64, 512) & 10,304 \\
MaxPool1D (size 2) & (64, 512) & (64, 256) & 0 \\
Conv1D + ReLU & (64, 256) & (128, 256) & 41,088 \\
MaxPool1D (size 2) & (128, 256) & (128, 128) & 0 \\
Flatten & (128, 128) & 16,384 & 0 \\
Dense 1 + ReLU & 16,384 & 256 & 4,194,560 \\
Dense 2 + ReLU & 256 & 128 & 32,896 \\
Output & 128 & 1 & 129 \\
\midrule
\textbf{Total} & & & \textbf{4,279,169} \\
\bottomrule
\end{tabular}
\end{table}

\subsection{Transformer}

\begin{table}[H]
\centering
\caption{Decoder-only Transformer architecture (autoregressive one-step
predictor; model dimension $128$, $4$ attention heads, causal mask,
post-layer normalization). Shapes are (sequence length, features), with the
batch dimension omitted.}
\label{tab:transformer}
\begin{tabular}{llll}
\toprule
\textbf{Layer} & \textbf{Input Shape} & \textbf{Output Shape} & \textbf{Parameters} \\
\midrule
Input projection & (1024, 1) & (1024, 128) & 256 \\
LayerNorm & (1024, 128) & (1024, 128) & 256 \\
Sinusoidal positional encoding & (1024, 128) & (1024, 128) & 0 \\
\multicolumn{4}{c}{\textit{Decoder block 1}} \\
\quad Masked self-attention & (1024, 128) & (1024, 128) & 66,048 \\
\quad Add \& LayerNorm & (1024, 128) & (1024, 128) & 256 \\
\quad FFN 1 + GELU & (1024, 128) & (1024, 512) & 66,048 \\
\quad FFN 2 & (1024, 512) & (1024, 128) & 65,664 \\
\quad Add \& LayerNorm & (1024, 128) & (1024, 128) & 256 \\
\multicolumn{4}{c}{\textit{Decoder block 2 (identical)}} \\
\quad Block total & (1024, 128) & (1024, 128) & 198,272 \\
Last-position readout & (1024, 128) & (1, 128) & 0 \\
Output linear 1 + GELU & (1, 128) & (1, 256) & 33,024 \\
Output linear 2 + GELU & (1, 256) & (1, 128) & 32,896 \\
Output linear 3 & (1, 128) & (1, 1) & 129 \\
\midrule
\textbf{Total} & & & \textbf{463,105} \\
\bottomrule
\end{tabular}
\end{table}

\subsection{Variational autoencoder}

\begin{table}[H]
\centering
\caption{VAE architecture (input length $1024$, latent dimension $32$;
batch dimension shown as $B$).}
\label{tab:vae}
\small
\setlength{\tabcolsep}{4pt}
\begin{tabular}{llll}
\toprule
\textbf{Component} & \textbf{Specification} & \textbf{Activation} & \textbf{Output Shape} \\
\midrule
\multicolumn{4}{c}{\textit{Encoder}} \\
Conv1D & In: 1, Out: 32, K: 4, S: 2, P: 1 & BatchNorm+LeakyReLU & ($B$, 32, 512) \\
Conv1D & In: 32, Out: 64, K: 4, S: 2, P: 1 & BatchNorm+LeakyReLU & ($B$, 64, 256) \\
Conv1D & In: 64, Out: 128, K: 4, S: 2, P: 1 & BatchNorm+LeakyReLU & ($B$, 128, 128) \\
Flatten & - & - & ($B$, 16384) \\
Linear ($\boldsymbol{\mu}$) & In: 16384, Out: 32 & - & ($B$, 32) \\
Linear ($\log \boldsymbol{\sigma}^2$) & In: 16384, Out: 32 & - & ($B$, 32) \\
\multicolumn{4}{c}{\textit{Latent Space}} \\
Reparameterize & $\mathbf{z} = \boldsymbol{\mu} + \boldsymbol{\epsilon} \odot \boldsymbol{\sigma}$ & - & ($B$, 32) \\
\multicolumn{4}{c}{\textit{Decoder}} \\
Linear & In: 32, Out: 16384 & - & ($B$, 16384) \\
Reshape & - & - & ($B$, 128, 128) \\
ConvTranspose1D & In: 128, Out: 64, K: 4, S: 2, P: 1 & BatchNorm+LeakyReLU & ($B$, 64, 256) \\
ConvTranspose1D & In: 64, Out: 32, K: 4, S: 2, P: 1 & BatchNorm+LeakyReLU & ($B$, 32, 512) \\
ConvTranspose1D & In: 32, Out: 1, K: 4, S: 2, P: 1 & - & ($B$, 1, 1024) \\
\midrule
\textbf{Total} & & & \textbf{1,672,449} \\
\bottomrule
\end{tabular}
\end{table}

\subsection{Denoising diffusion probabilistic model}

The noise predictor is a one-dimensional U-Net (Table~\ref{tab:ddpm}). Each
block applies two convolutions of kernel size $3$, each followed by ReLU and
batch normalization, adds a learned projection of the timestep embedding
between them, and ends with a stride-$2$ convolution (encoder) or transposed
convolution (decoder) that halves or doubles the length. Decoder blocks take
the concatenation of the previous output and the skip connection from the
encoder block at the same resolution. The timestep is encoded by a
$64$-dimensional sinusoidal embedding followed by a linear layer and ReLU.
The five- and four-layer variants of Section~5.3 of the main text keep the
first five or four channel widths ($32$ to $512$ or $32$ to $256$) and have
$5{,}788{,}209$ and $1{,}471{,}537$ parameters, respectively.

\begin{table}[H]
\centering
\caption{DDPM 1D U-Net architecture (default configuration: $6$ down/up
blocks, input length $1024$; Section~5.3 of the main text varies the input
length and depth).}
\label{tab:ddpm}
\small
\begin{tabular}{lll}
\toprule
\textbf{Stage} & \textbf{Block} & \textbf{Specification} \\
\midrule
Input & Time series & Length: 1024, Channels: 1 \\
\midrule
\multicolumn{3}{c}{\textit{Encoder (down-sampling)}} \\
Block 1 & Conv1D & Channels: $1 \to 32$, Length: $1024 \to 512$ \\
Block 2 & Conv1D & Channels: $32 \to 64$, Length: $512 \to 256$ \\
Block 3 & Conv1D & Channels: $64 \to 128$, Length: $256 \to 128$ \\
Block 4 & Conv1D & Channels: $128 \to 256$, Length: $128 \to 64$ \\
Block 5 & Conv1D & Channels: $256 \to 512$, Length: $64 \to 32$ \\
Block 6 & Conv1D & Channels: $512 \to 1024$, Length: $32 \to 16$ \\
\midrule
\multicolumn{3}{c}{\textit{Bottleneck}} \\
Attention & Self-attention & 4 heads, Length: 16, Channels: 1024 \\
\midrule
\multicolumn{3}{c}{\textit{Decoder (up-sampling)}} \\
Block 1 & ConvTranspose1D + Skip & Channels: $1024 \to 512$, Length: $16 \to 32$ \\
Block 2 & ConvTranspose1D + Skip & Channels: $512 \to 256$, Length: $32 \to 64$ \\
Block 3 & ConvTranspose1D + Skip & Channels: $256 \to 128$, Length: $64 \to 128$ \\
Block 4 & ConvTranspose1D + Skip & Channels: $128 \to 64$, Length: $128 \to 256$ \\
Block 5 & ConvTranspose1D + Skip & Channels: $64 \to 32$, Length: $256 \to 512$ \\
Block 6 & ConvTranspose1D + Skip & Channels: $32 \to 16$, Length: $512 \to 1024$ \\
Output & Conv1D & Channels: $16 \to 1$, Length: $1024$ \\
\midrule
Diffusion & Timesteps & $T = 1000$, linear schedule: $\beta_1 = 10^{-4}, \beta_T = 0.02$ \\
\midrule
\textbf{Total (6 layers)} & & \textbf{22,941,233 parameters} \\
\bottomrule
\end{tabular}
\end{table}

\section{Training Protocol}

\paragraph{Training data.}
All four architectures are trained on a single trajectory of $10^{5}$ points
(for ETG, $0.1$~ns sampled every $1$~fs). A second trajectory is not used
for training: it supplies the initial windows of the autoregressive rollouts,
and for the CNN and the Transformer a loss on it is recorded during training
for monitoring only. It plays no role in model selection, early stopping, or
the learning-rate schedule, and the VAE and the DDPM, which generate
unconditionally, do not use it at all; no validation set is used for any
model. The training trajectory is normalized to zero mean and unit variance and cut
into $1024$-point windows with stride $10$, giving $9{,}898$ training windows
(Eq.~(27) of the main text); for the autoregressive models the point following
each window is the prediction target.

\paragraph{Reference data.}
The converged real-data reference used throughout the main text is built from
$100$ statistically independent trajectories per system ($0.1$~ns each,
$10$~ns in total), drawn from the same source as the training data: for OU,
realizations generated with the same integration algorithm and parameters;
for ETG, MD trajectories from the same simulation campaign as the training
trajectory, with $100$ trajectories available for each of the $15$
conformations, so that each conformation in Section~5.4 of the main text is
scored against its own $100$-trajectory reference. The training trajectory
and the second trajectory described above are among these $100$, so the
training data make up $1\%$ of the reference data. The real-data curve of
the convergence study (Section~5.5 of the main text) uses a further $96$ MD
trajectories of conformation $15$, disjoint from both the training and the
reference data.

\paragraph{Optimization.}
Training hyperparameters are listed in Table~\ref{tab:training}. The VAE and
the DDPM use the Adam optimizer~\cite{Kingma2014}, and the CNN and the
Transformer its weight-decay variant AdamW. All four
models clip gradients to a norm of $1.0$. The CNN and Transformer minimize
the squared one-step prediction error and use a one-cycle learning-rate
schedule (cosine annealing, $30\%$ warm-up). The VAE minimizes the summed
squared reconstruction error plus the Kullback--Leibler term weighted by
$10^{-3}$. The DDPM minimizes the noise-prediction error of Eq.~(32) of the
main text, with noising steps drawn uniformly from the $N_{\mathrm{d}}=1000$ steps of the
linear noise schedule ($\beta_j$ from $10^{-4}$ to $0.02$). For the VAE and the DDPM the learning rate is halved
whenever the epoch-averaged training loss has not improved for $5$ epochs.

\begin{table}[H]
\centering
\caption{Training hyperparameters of the reported configurations.
$\lambda$: weight decay. One-cycle: cosine one-cycle schedule with $30\%$
warm-up, peak learning rate listed. Plateau: learning rate halved after $5$
epochs without improvement of the training loss. Dropout is $0$ for the CNN
and the Transformer; the VAE latent dimension is $32$.}
\label{tab:training}
\begin{tabular}{llllll}
\toprule
\textbf{Model} & \textbf{Optimizer} & \textbf{Learning rate} & \textbf{Schedule} & \textbf{Batch} & \textbf{Epochs} \\
\midrule
CNN & AdamW ($\lambda=10^{-5}$) & $10^{-3}$ & one-cycle & 32 & 50 \\
Transformer & AdamW ($\lambda=10^{-2}$) & $3\times10^{-4}$ & one-cycle & 32 & 50 \\
VAE & Adam & $10^{-3}$ & plateau & 64 & 100 \\
DDPM & Adam & $10^{-3}$ & plateau & 64 & 100 \\
\bottomrule
\end{tabular}
\end{table}

\paragraph{Generation.}
The autoregressive models generate each trajectory by a $1024$-step rollout
from a window drawn at random from the second trajectory, feeding each
prediction back as input; rollouts whose matched true continuation would run
past the end of that trajectory (about $1\%$) are discarded. The VAE decodes
latent vectors drawn from the standard normal prior, and each decoded window
is rescaled to zero mean and unit variance before denormalization, because
the decoder output is under-dispersed. The DDPM runs the reverse diffusion
process from Gaussian noise (Section~\ref{si:instability_sub}). All
generated trajectories are denormalized with the mean and standard
deviation of the training trajectory (for the autoregressive models, of the
trajectory that supplies the initial windows).

\paragraph{Hyperparameter selection.}
For each model a set of configurations was trained on the same data and
the reported one was retained. For the CNN and the Transformer, the
architecture and network size were tuned in preliminary runs on the same
training data; among the final configurations, batch sizes of $32$ and $64$,
$50$ and $100$ epochs, and dropout of $0.1$ and $0$ were compared. A window stride of $1$ was also tested for both; the stride
was nevertheless kept at $10$ for all four models, so that every model sees
the same training windows. For the VAE, batch sizes of $32$ and $64$ and
latent dimensions of $32$, $64$, $128$, and $256$ were compared
($1.7\times10^{6}$ to $1.3\times10^{7}$ parameters). For the DDPM, batch
sizes of $32$ and $64$ were compared, and the input length, window stride,
and U-Net depth were varied about the default as reported in Section~5.3 of
the main text.

\section{Sampling Stability and Reproducibility}
\label{si:instability}

\subsection{Rare divergent draws}
\label{si:instability_sub}

Individual draws occasionally diverge during sampling even though training
converges normally (final loss $\approx0.018$). The cause is the U-Net's
batch-normalization layers, whose running variance reaches
$10^{5}$--$10^{7}$, so that evaluation-mode sampling has a small probability
per draw of entering a runaway feedback loop that grows by $\sim\!6\times$
per denoising step until it overflows. This is a known fragility of batch
normalization in diffusion U-Nets that would be removed by switching to
group normalization. Divergent draws are rare: most runs produce none, and
no run used in this work produced more than $5$ in $1000$. Divergent draws,
identified as non-finite or as having a standard deviation more than ten
times the ensemble median, are discarded and redrawn from the same sampler,
and the count is recorded for every run. The remaining samples show no
systematic difference from those of runs that produce no divergent draw. Because the failure is a
numerical divergence of the reverse process rather than a property of the
learned distribution, and because a single such draw would otherwise
corrupt every ensemble average, this is a sampling-time safeguard rather
than a change of model: on runs that produce no divergent draw, it leaves
the output bit-for-bit identical.

\subsection{Sampling noise and the median over generation draws}
\label{si:sampling_noise}

The diffusion hyperparameter sweep and the conformation study (Sections~5.3
and~5.4 of the main text) report, for each configuration, the median over
five independent generation draws. Each draw is a fresh ensemble of $1000$
trajectories sampled from the same trained network with a different
sampling seed; the network is not retrained. This section documents the
scatter that motivates this choice.

For a fixed trained network the draw-to-draw scatter is large. For the
default configuration, the five draws give MSE\,$\log_{10}$PSD between
$3.8\times10^{-4}$ and $1.7\times10^{-3}$, a $4.6$-fold spread, and three
further independent trainings of the same configuration give spreads of
$2.4$, $9.0$, and $5.3$-fold. Across the configurations of the
hyperparameter sweep, the spread ranges from $1.3$-fold (stride $2$) to
$31$-fold (stride $5$, due to a single outlying draw).

Training-to-training variance is smaller. The default configuration was
retrained from scratch three times on the same training trajectory with
different seeds (training on the
GPU is not bit-reproducible, so each training is an independent
realization). Summarized by the median of its five draws, each of the four
trainings gives MSE\,$\log_{10}$PSD between $7.0\times10^{-4}$ and
$1.3\times10^{-3}$, a $1.8$-fold spread, smaller than the draw-to-draw spread
of a single network. Generation noise, rather than training variance, is
therefore the dominant source of run-to-run scatter. Because it is
generation noise, it can be averaged cheaply: no retraining is required,
only repeated sampling from an unchanged network, which is why the median
of five draws is reported for every configuration.

\section{Error Bars in the Convergence Study}
\label{si:convergence_errbars}

This section defines the points and error bars of Fig.~7 of the main text
(Section~5.5). For each sample size $N$, $400$ random subsets of $N$ windows
are drawn without replacement from a pool. For synthetic data the pool is
$4\times10^{4}$ samples generated by the default diffusion configuration
(training seed $3$ of four, the middle of the range; seed $0$ is used
elsewhere in the paper). For real data it is the non-overlapping windows of
$96$ MD trajectories disjoint from the training and reference data, which
limits the real series to $N\le4000$ ($t\le4.1$~ns). At $N=4\times10^{4}$ the
subset is the whole synthetic pool, so a single evaluation is used.

Each subset is split into ten groups of $N/10$ windows, and each statistic is
averaged within each group; $J(\omega)$ is computed from each group's mean
$C_2(\tau)$. The point is the MSE of the mean over the ten groups with
respect to the converged reference, averaged over the lag or frequency grid.
The error bar is the squared standard deviation of the ten group averages,
averaged over the same grid; for $\log_{10}$PSD the standard deviation
$\sigma$ is first propagated to $\log_{10}$ as $\sigma/(\ln 10\,\bar{P})$,
with $\bar{P}$ the mean PSD. Both the point and the error bar are the medians
of these quantities over the $400$ subsets, so the error bar has the units of
the MSE it accompanies. It measures the scatter of a statistic between groups
of $N/10$ windows, not the uncertainty of the plotted median. The median is
used because a single subset is unrepresentative: for real data at the
training-set size, the central $68\%$ of single-subset values of
MSE\,$\log_{10}$PSD spans a factor of about $20$.

\bibliography{refs}